\documentclass[fleqn,usenatbib]{mnras}

\usepackage[T1]{fontenc}
\usepackage{ae,aecompl}

\usepackage{graphicx}	
\usepackage{amsmath}	
\usepackage{amssymb}	
\usepackage{subfigure}

\newcommand{\RR}{{\cal R}}

\newcommand{\AAA}{{\cal A}}

\newcommand{\al}{\alpha}
\newcommand{\bal}{{\boldsymbol\alpha}}

\newcommand{\de}{\delta}
\newcommand{\De}{\Delta}
\newcommand{\ep}{\epsilon}
\newcommand{\bep}{{\boldsymbol\epsilon}}
\newcommand{\gga}{\gamma}

\newcommand{\ka}{\kappa}

\newcommand{\La}{\Lambda}
\newcommand{\Om}{\Omega}
\newcommand{\om}{\omega}
\newcommand{\si}{\sigma}
\newcommand{\Si}{\Sigma}

\newcommand{\bth}{{\boldsymbol\theta}}
\newcommand{\ra}{\rightarrow}

\newcommand{\bnabla}{{\boldsymbol\nabla}}
\newcommand{\be}{\begin{equation}}
\newcommand{\ee}{\end{equation}}

\newcommand{\bea}{\begin{eqnarray}}
\newcommand{\eea}{\end{eqnarray}}
\newcommand{\bean}{\begin{eqnarray*}}
\newcommand{\eean}{\end{eqnarray*}}
\newcommand{\dd}{\partial}

\newcommand{\ie}{{\em i.e. }}
\newcommand{\eg}{{\em e.g. }}
\newcommand{\ii}{\mathrm{i}}
\newcommand{\er}{\mathrm{e}}
\newcommand{\id}{{\rm 1\kern -2.5pt I}} 
\newcommand{\bn}{{\mathbf n}}
\newcommand{\bde}{{\mathbf e}}
\def\mean#1{\left< #1 \right>}

\title[Weak-lensing observables in relativistic simulations]{Weak-lensing observables in relativistic $N$-body simulations}

\author[Lepori, Adamek, Durrer, Clarkson \& Coates]{Francesca Lepori,$^{1}$\thanks{francesca.lepori@unige.ch}
Julian Adamek,$^{2}$\thanks{\!julian.adamek@qmul.ac.uk}
Ruth Durrer,$^{1}$\thanks{ruth.durrer@unige.ch}
Chris Clarkson$^{2}$\thanks{chris.clarkson@qmul.ac.uk}
\newauthor and Louis Coates$^{2}$\thanks{l.j.c.coates@qmul.ac.uk}
\\
$^{1}$D\'epartement de Physique Th\'eorique and Centre for Astroparticle Physics, Universit\'e de Gen\`eve\\
~~24 quai Ernest-Ansermet,  CH-1211 Gen\`eve 4, Switzerland \\
$^{2}$School of Physics \& Astronomy, Queen Mary University of London, 327 Mile End Road, London E1 4NS, United Kingdom
}

\date{Accepted XXX. Received YYY; in original form ZZZ}

\pubyear{2019}

\begin{document}
\label{firstpage}
\pagerange{\pageref{firstpage}--\pageref{lastpage}}
\maketitle

\begin{abstract}
We present a numerical weak-lensing analysis that is fully relativistic and non-perturbative for the scalar part of the gravitational potential and first-order in the vector part, frame dragging. Integrating the photon geodesics backwards from the observer to the emitters, we solve the Sachs optical equations and study in detail
the weak-lensing convergence, ellipticity and rotation.
For the first time, we apply such an analysis to
a high-resolution relativistic $N$-body simulation, which consistently includes 
the leading-order corrections due to general relativity on both large and small scales. These are related to the question of gauge choice and to post-Newtonian corrections, respectively. 
We present the angular power spectra and one-point probability distribution functions for the weak-lensing variables, which we find are broadly in agreement with comparable Newtonian simulations. Our geometric approach, however, is more robust and flexible, and can therefore be applied consistently to non-standard cosmologies and modified theories of gravity.
\end{abstract}




\section{Introduction}

In the near future galaxy surveys like DESI, LSST, Euclid and others, will map out nearly the entire visible Universe~-- see~\citet{Abate:2012za,Abell:2009aa,Aghamousa:2016zmz,Amendola:2016saw,Santos:2015bsa,4MOST:2019}. They will measure redshifts, angular positions but also shapes and sizes of billions of galaxies with unprecedented precision.
Because the weak gravitational lensing signal in particular is dominated by nonlinear structures, linear perturbation theory will not be sufficient to interpret these data sets, and numerical simulations are a vital tool which analysis pipelines will have to rely on.

Such simulations are commonly based on Newtonian gravity. However, 
recently some of us have proposed a new $N$-body code based on general relativity~\citep{Adamek:2015eda,Adamek:2016zes}. In this code, all relativistic effects which can be relevant on scales much larger than the biggest black hole horizon are taken into account. Metric perturbations are included at first order only, but their spatial derivatives are kept at all orders. This guarantees a consistent treatment of the gauge issue on cosmological scales as well as to capture the first post-Newtonian order at small scales, see \citet{Adamek:2015eda,Adamek:2016zes} for further details.

In the past, structure formation in $\La$CDM and with massive neutrinos have been studied with this code~\citep{Adamek:2014xba,Adamek:2017uiq}, and it has been extended to simulate clustering dark energy \citep{Hassani:2019lmy} as well as $f(R)$ gravity \citep{Reverberi:2019bov}. Recently some of us have also studied fully relativistic photon geodesics in this code to model the distance--redshift relation including clustering~\citep{Adamek:2018rru}. The present work builds upon this development.

Here we want to study weak lensing. We compute the relativistic ellipticity $\ep$ (an observable closely related to the weak-lensing shear), the convergence $\ka$ as well as the rotation $\om$ of images due to foreground structure non-perturbatively from a high-resolution simulation. We determine their angular power spectra for different redshifts as well at their one-point probability distribution functions. Because our analysis is non-perturbative, we can study effects like rotation, which is a purely non-linear effect which is absent within linear perturbation theory. A recent study has found \citep{Deshpande:2019sdl} that a linear analysis will also be insufficient for interpreting the weak-lensing shear signal in upcoming surveys.

Following~\citet{Perlick:2010zh}, we develop a fully relativistic geometric description of lensing which is valid in arbitrary spacetimes, but we also make contact with the more familiar standard treatment in cosmological perturbation theory. This is the first study of weak lensing which uses a relativistic $N$-body simulation of ``production scale'' -- \ie with multi-billion mass elements. Relativistic simulations of weak lensing have been done in the past for small problem sizes, see \citet{Giblin:2017ezj} for an example, but these do not probe the small-scale structure that is responsible for the bulk of the lensing signal. We find that our results agree very well with previous Newtonian and perturbative approaches within the present context of the standard $\La$CDM cosmological model.

The remainder of this paper is organised as follows:
In the next section we describe our methods. After a brief introduction to the general theory of weak lensing, we present the non-perturbative optical equations in Poisson gauge which are solved numerically in this work. In Section~3 we briefly describe our simulation and  Section~4 is devoted to results. We present sky maps of the relevant lensing signals as well as angular power spectra and probability distribution functions. In Section~5 we conclude. Some more technical issues are relegated to four appendices.

\section{Method}

\subsection{Weak-lensing theory}
\label{sec:theory}

When describing the properties of an infinitesimal bundle of light rays it is useful to introduce a ``screen'' that, at any given point along the central ray, is defined by two spacelike orthonormal screen vectors $e_1^\mu$, $e_2^\mu$ that are normal to the photon four-vector $k^\mu$. Each choice of such a screen basis corresponds to a particular reference frame with four-velocity $u^\mu$ that is normal to the screen vectors. However, one can show that the shape and size of an \textit{infinitesimal} beam cross-section does not depend on this choice \citep{Sachs:1961zz}. In other words, the proper motion of the screen has no effect on an infinitesimal image that is projected onto it.

One particularly useful choice is then a screen basis that is parallel-transported along the central ray. Such a choice is called a Sachs basis. It draws a straightforward connection between the orientation of an image at an observer location and the aspect of the source in the screen basis. In cosmological spacetimes with a preferred global timelike vector field $u^\mu$ one sometimes chooses a screen basis that is everywhere orthogonal to $u^\mu$, but in this case the screen is in general not parallel-transported. This choice can still be useful, \eg in perturbative calculations where $u^\mu$ can be chosen as the matter rest frame, see \citet{Pitrou:2012ge,Marozzi:2016qxl} for examples\footnote{Note that \citet{Marozzi:2016qxl} and some other authors use the term ``Sachs basis'' for such a screen basis if its projection into the screen space is parallel-transported, which can be arranged for any choice of $u^\mu$. This differs from our convention that follows \citet{Sachs:1961zz,Perlick:2010zh}, and the two conventions coincide if and only if $u^\mu$ itself is parallel-transported.}. However, a unique notion of matter rest frame is not always available in the real Universe, due to the existence of voids and overlapping matter streams. While we will later choose an observer at rest with the CMB (which defines a preferred frame $u^\mu$) we shall discuss lensing purely from a spacetime perspective and work with a Sachs basis.

A small source mapped into a small image can be considered as a linear map from the screen at the source to the screen at the observer. The Jacobi map, $D$, describes the deformation of such a small image in the Sachs basis. In this treatment we follow~\citet{Perlick:2010zh}.
We first introduce the deformation matrix $S$ of a light bundle as
\be
S = \left(\begin{array}{cc} \theta +\si_1 & \si_2 \\ \si_2 &  \theta -\si_1\end{array}\right)
\ee
Here we use the geometric optics approximation within which the photon wave vector is the gradient of the eikonal so that $S_{AB}= e_A^\mu e_B^\nu k_{\mu;\nu}$ is symmetric. The deformation matrix is related to spacetime curvature through
\be
\dot S + S^2 = \RR S\,,
\ee
where a dot denotes the derivative with respect to the affine parameter of the photon geodesic and
$\RR$ is given by contractions of the Ricci and Weyl tensors with the photon four-vector
and the so-called screen vectors (the aforementioned Sachs basis), see~\cite{Perlick:2010zh} for more details. Written out in components and introducing the complex null shear $\si \equiv \si_1 + \ii \si_2$, one finds
\be\label{e:thdot}
    \dot{\theta} + \theta^2 + \si \si^\ast = -\frac{1}{2} R_{\mu\nu} k^\mu k^\nu\,,
\ee
\be\label{e:sidot}
    \dot{\si} + 2 \theta \si = -\frac{1}{2} C_{\alpha\mu\beta\nu} k^\mu k^\nu \left(e_1^\alpha + \ii e_2^\alpha\right) \left(e_1^\beta + \ii e_2^\beta\right)\,,
\ee
where $R_{\mu\nu}$ and $C_{\alpha\mu\beta\nu}$ are the Ricci and Weyl curvature tensors, respectively. The Jacobi map $D$ then satisfies 
\be\label{e:dotD}
\dot D = DS \,.
\ee

In complete generality, we can parametrize $D$ as
\be\label{e:Dpar}
 D = D_AR(-\chi-\om) \left(\begin{array}{cc} \er^\gga & 0 \\ 0 &  \er^{-\gga}\end{array}\right)R(\chi) \,.
\ee
Here $R(\al)$ denotes the rotation by an angle $\al$,
\be
R(\al)= \left(\begin{array}{cc} \cos\al & \sin\al \\ -\sin\al &  \cos\al \end{array}\right) \,.
\ee
With this choice of parametrisation $D_A=\sqrt{\det D}=\sqrt{D_+D_-}$ is the area distance, where $D_\pm = D_A\exp(\pm\gga)$ are the eigenvalues of $R(\om)D=\Si$ which is symmetric. The rotation by the angle $\chi$ rotates $\Si$ into its eigen directions.

We could also have placed the rotation by the angle $\om$ on the other side:
for an arbitrary $2\times2$ matrix $M$ there exists and angle $\om$ such that $R(\om)M=\Si$ is symmetric. In this case also  $MR(\om)=\Si'$ is symmetric and
$\Si' = R(-\om)\Si R(\om)$.
The proof of this proposition is very simple and can be done by construction. 
In our application this would correspond in replacing $\chi$ by $\chi+\om$.

It is convenient to introduce the complex ellipticity $\ep$,
\be
\ep \equiv \left(\frac{D_+}{D_-}-\frac{D_-}{D_+}\right)\er^{2\ii\chi} = 2 \er^{2\ii\chi} \sinh2\gga \,.
\ee
Its phase encodes $\chi$ and its modulus is the product of the first and second eccentricity of the image of a circle under the Jacobi map. It is therefore closely related to the shear of the image (parametrised by $\gga_1$ and $\gga_2$) as we will see later.

The differential equation \eqref{e:dotD} can be written as [see~\citet{Perlick:2010zh}, eq. (27)]
\be
\dot D_\pm +\ii\dot\chi D_\pm -\ii(\dot\chi+\dot\om)D_\mp = (\theta \pm \er^{-2\ii\chi}\si)D_{\pm} \,,
\ee
which can be rearranged into following equations for $D_A$, $\ep$ and $\om$:
\be
\label{e:DAdot}
\dot D_A = \theta D_A \,,
\ee
\be
\label{e:epdot}
\dot\ep = 4\si \cosh2\gga = 2 \si \sqrt{4+\ep\ep^\ast} \,,
\ee
\be
\label{e:omdot}
\dot\om = (\si_1\sin2\chi -\si_2\cos2\chi) \tanh\gga = \frac{\ii\left(\ep^\ast \si - \ep \si^\ast\right)}{4 + 2 \sqrt{4+\ep\ep^\ast}}\,.
\ee

It is interesting to note that the image rotation $\om$, which quantifies the net rotation with respect to a parallel-transported screen basis, is in full generality small at second order. This has also been found in \citet{DiDio:2019rfy} where an analysis for scalar, vector and tensor perturbations was presented.

The boundary condition for $D$ is $D(s_0)=0$ at the observer, which implies that
all the quantities $D_A$, $\si$, $\ep$ and $\om$ can be set to zero in the initial (or rather final) conditions. Note also that $\chi$ is not a perturbative parameter. While $\gga$ and $\om$ are expected to remain small, $\chi$ is simply the rotation into the principal axes which can be a large rotation even if the deformation of the light bundle is arbitrarily small.
For this reason it is simpler to work with the complex ellipticity $\ep$ instead of $\gga$ and $\chi$.
If $\ep=0$ the ellipse described by the symmetric matrix $\Si= R(\om)D$ becomes a circle and the angle $\chi$ is ill defined.
The evolution equation (\ref{e:epdot}) of $\ep$, however, remains well-behaved at such singular points.

In the literature on weak gravitational lensing
the Jacobi map is often written as [see \eg \cite{RuthBook}, eq.~(7.18)]
\be\label{e:AAA}
 D = \bar D_A \AAA \,, \qquad \AAA =  \left(\!\begin{array}{cc} 1-\ka_{\rm{lin}}-\gga_1 & \nu -\gga_2 \\ -\nu -\gga_2 & 1-\ka_{\rm{lin}}+ \gga_1\end{array}\!\right)
 \ee
where $\bar D_A$ is the area distance of the background Friedmann metric and $\AAA$ is called the amplification matrix. The trace of $\AAA$ is parametrised by the convergence $\ka_{\rm{lin}}$, and the shear of the image is given by $\gga_1$, $\gga_2$. Usually one sets $\nu=0$. At first order, these quantities are 
related to the area distance, the complex ellipticity and the rotation via
\be
\kappa \equiv 1- \frac{D_A}{\bar{D}_A}\simeq \ka_{\rm{lin}}\,,
\qquad \gga_1 + \ii \gga_2 \simeq -\frac{ \ep}{4}\,,  \qquad \om\simeq\nu \,.
\label{conv-shear}
\ee
Even though $\AAA$ is the most general ansatz for a $2\times2$ matrix, we prefer to work with \eqref{e:Dpar} since the interpretation of its elements is straightforward:
$\mu=\bar D_A^2/D_A^2$ is the magnification, $\gga$ is the amplitude of the shear and $R(\chi)$ determines its principal axes, and finally $\om$ determines the rotation of the image. Note that already at second order in the parametrisation \eqref{e:AAA} the quantities are mixed, for example the magnification becomes
\be
\mu\, \equiv \det \AAA^{-1} = \frac{1}{\left(1-\kappa\right)^2} = \frac{1}{\left(1-\kappa_{\rm{lin}}\right)^2 - \gamma_1^2 - \gamma^2_2+\nu^2}\,.
\label{magn}
\ee
The fully non-perturbative expressions for $\gga_1+\ii\gga_2$ and $\nu$ are
\be
\gga_1+\ii\gga_2 = -\frac{D_A}{\bar D_A}\er^{\ii(2\chi+\om)}\sinh\gga
= -\frac{\left(1-\ka\right) \ep \er^{\ii\om}}{2\sqrt{2+\sqrt{4+\ep\ep^\ast}}}\,, 
\ee
\be
\nu =\frac{D_A}{\bar D_A}\sin\om \cosh\gga 
= \frac{1-\ka}{2}\sin\om\sqrt{2+\sqrt{4+\ep\ep^\ast}}\,.
\ee

Therefore, at next-to-leading order, both the trace-free symmetric part of $\AAA$ (parametrised by $\gga_1 + \ii \gga_2$) and the antisymmetric part of $\AAA$ (given by $\nu$) are actually mixtures of image rotation and shear.
Since the quantities $\ka,~\gga_1,~\gga_2$ and $\nu$ have simple geometrical interpretations only at first order in perturbation theory, we shall work with the fully non-perturbative quantities $D_A$ or $D_A/\bar D_A$, $\ep$ and $\om$.

As we discuss in detail in Section~\ref{ss:optical} and \ref{ss:ray}, in our numerical approach we compute $D_A$, $\ep$ and $\om$ without employing any approximations in the scalar sector, and taking into account also the leading-order corrections due to general-relativistic frame dragging.

\subsection{Optical equations in Poisson gauge}\label{ss:optical}

Up to this point our discussion was completely general in the sense that we did not have to specify the spacetime metric. We will now specialise to the case of a perturbed Friedmann-Lema\^itre-Robertson-Walker (FLRW) metric, where the line element is written in Poisson gauge as
\be
ds^2 = a^2(\tau)\! \left[-\er^{2\psi} d\tau^2 \!-\! 2 B_i dx^i d\tau \!+\! \left(\er^{-2\phi} \delta_{ij} \!+\! h_{ij}\right)\!dx^i dx^j\right].
\ee
Here, $\tau$ is conformal time and we assume a spatially flat background with scale factor $a(\tau)$ and comoving Cartesian coordinates $x^i$. In $\Lambda$CDM cosmology the metric perturbations are dominated by the two scalar gravitational potentials $\psi$ and $\phi$, which we will treat non-perturbatively, although they do remain small, of the order of $10^{-5}$ at the scales that we probe. The divergence-free vector potential $B_i$, which captures the frame-dragging effect, is at least two orders of magnitude smaller than the scalar potentials \citep{Lu:2008ju,Thomas:2015kua,Adamek:2015eda} and will therefore be treated at leading order only. This means we will neglect terms that are quadratic in $B_i$ and also approximate $\er^\phi B_i \simeq \er^\psi B_i \simeq B_i$. A comparison with numerical relativity simulations \citep{Adamek:2020jmr} demonstrates that this is an extremely good approximation. The spin-2 field $h_{ij}$, which is transverse and traceless, is even smaller on the scales that we are interested in, and we will therefore neglect it here. It would not pose a conceptual challenge to include it, but it would make our equations more cumbersome.

For a classical point particle, let us introduce the canonical peculiar momentum per unit
mass as
\be
 q_i \equiv m^{-1}_p \frac{\partial\mathcal{L}}{\partial\frac{dx^i_p}{d\tau}}\,,
\ee
where $m_p$ is the mass of the particle and $\mathcal{L}$ is the Lagrangian describing the geodesic motion of its comoving coordinate $x^i_p$. The particle's four-velocity vector is
\begin{multline}
 u^\mu = \delta^{\mu0} \frac{\er^{-\psi}}{a^2} \sqrt{q^2 \er^{2\phi} + a^2}
 \\  + \delta^{\mu i} \left(\!\frac{q_i \er^{2\phi}}{a^2} + \frac{B_i}{a^2}\sqrt{q^2 \er^{2\phi} + a^2} \!\right),
\end{multline}
where
$ q^2 =\de^{ij}q_iq_j$.

A local Fermi frame is given by a spatial triad of orthonormal basis vectors $s^\mu_{(i)}$ that are orthogonal to $u^\mu$, \ie $g_{\mu\nu} u^\mu s^\nu_{(i)} = 0$ and $g_{\mu\nu}s^\mu_{(i)}s^\nu_{(j)} = \delta_{ij}$. For $q = 0$ (the cosmological rest frame) we can choose $s^\mu_{(i)} = \delta^\mu_i a^{-1} \er^{\phi}$.

For the tangent vector of a photon geodesic, $k^\mu$, the null condition implies
\be\label{e:nullcondition}
\frac{k^i}{k^0} = \er^{\phi+\psi} n^i + \delta^{ij} B_j\,,
\ee
where $n^i$ is normalised as $n^i n^j \delta_{ij} = 1$ and denotes the direction of the photon path in the cosmological rest frame,
\be
n^i = -\delta^{ij}\left.\frac{g_{\mu\nu} k^\mu s^\nu_{(j)}}{g_{\alpha\beta} k^\alpha u^\beta}\right\vert_{q=0}\,.
\ee

Since we are interested in the properties of null geo\-desics it is convenient to employ a few conformal transformations in order to arrive at simple equations. First, we define $\tilde{k}^0 \equiv a^2 \er^{-2\phi} k^0$, such that the geodesic equations for the photon can be rewritten in terms of conformal time as
\be\label{e:photk0}
\frac{d\ln \tilde{k}^0}{d\tau} + 2 n^i \partial_i \er^{\phi+\psi} + \phi' + \psi' + n^i n^j \partial_i B_j = 0\,,
\ee
\be\label{e:photki}
\frac{dn^i}{d\tau} - \left(n^i n^j - \delta^{ij}\right) \left[\partial_j \er^{\phi+\psi} + n^k \partial_j B_k\right] = 0\,,
\ee
where a prime denotes partial derivative with respect to conformal time and $\partial_i$ denotes a partial derivative with respect to $x^i$. In order to also cast the geodesic deviation equations, \eqref{e:thdot} and (\ref{e:DAdot})--(\ref{e:omdot}) into a convenient form, we define the rescaled quantities
\be\label{e:confDA}
 \tilde{D}_A \equiv D_A \frac{\er^\phi}{a}\,, \qquad \tilde{\si} \equiv \frac{\si}{k^0} \tilde{D}_A^2\,,
\ee
as in \citet{Adamek:2018rru}. $\tilde{D}_A$ then evolves according to
\begin{multline}\label{e:SachsDA}
\frac{d^2 \tilde{D}_A}{d\tau^2} + \frac{d\ln\tilde{k}^0}{d\tau} \frac{d\tilde{D}_A}{d\tau} + \biggl[-\frac{1}{2} \left(n^i n^j - \delta^{ij}\right) \er^{\phi+\psi} \partial_i \partial_j \er^{\phi+\psi}\biggr.\\ \biggl.+ \frac{1}{2} n^i n^j \partial_i B_j' + \frac{1}{2} n^i \partial^2 B_i\biggr] \tilde{D}_A + \frac{\tilde{\si}\tilde{\si}^\ast}{\tilde{D}_A^3} = 0\,.
\end{multline}

In order to solve the evolution of $\tilde{\si}$ we first need to construct a screen space, given by a pair of screen vectors $e^\mu_A$ ($A=1,2$) with $g_{\mu\nu} e^\mu_A e^\nu_B = \delta_{AB}$ that are also orthogonal to $k^\mu$ and are parallel-transported along the ray. We can write them in terms of two spacelike directions $\tilde{e}^i_A$ with $\delta_{ij} n^i \tilde{e}^j_A = 0$ and $\delta_{ij} \tilde{e}^i_A \tilde{e}^j_B = \delta_{AB}$ as
\be
 e^\mu_A = \delta^\mu_i \frac{\er^\phi}{a} \tilde{e}^i_A + \frac{\tilde{\beta}_A}{a^2} \er^{-2\psi} \frac{k^\mu}{k^0}\,,
\ee
where the $\tilde{\beta}_A$ are related to the timelike direction to which the screen is orthogonal. For an observer in the cosmological rest frame we have $\tilde{\beta}_A = 0$ as a boundary condition, and $\tilde{\beta}_A$ will remain small (of the order of a metric perturbation) along the ray. The parallel transport of the screen basis $e_A^\mu$ implies
\be\label{e:screenpt}
\frac{d\tilde{e}^i_A}{d\tau} - n^i \tilde{e}^j_A \left(\partial_j \er^{\phi+\psi} + n^k \partial_{(j} B_{k)}\right) - \tilde{e}^j_A \delta^{ik} \partial_{[j} B_{k]} = 0\,,
\ee
\be
\frac{d\tilde{\beta}_A}{d\tau} - \left(\phi'+\psi'\right)\tilde{\beta}_A + a \left(\er^{\phi+\psi} \tilde{e}^i_A \partial_i \er^\psi + n^i \tilde{e}^j_A \partial_{(i}B_{j)}\right) = 0\,,
\ee
where we give the second equation here only for completeness as it will not be needed. This is because $\tilde{\beta}_A$ would only appear multiplied by terms of order $B_i$ which can then be dropped.

The projection of the Weyl tensor onto the screen basis determines the evolution of $\tilde{\si}$ as
\begin{multline}\label{e:Sachssi}
\frac{d\tilde{\si}}{d\tau} + \frac{d\ln\tilde{k}^0}{d\tau} \tilde{\si} + \frac{\tilde{D}_A^2}{2} \left(\tilde{e}^i_1 \tilde{e}^j_1 - \tilde{e}^i_2 \tilde{e}^j_2 + \ii \tilde{e}^i_1 \tilde{e}^j_2 + \ii \tilde{e}^i_2 \tilde{e}^j_1\right) \\
\times \left[\er^{\phi+\psi}\partial_i\partial_j\left(\er^{\phi+\psi} + n^k B_k\right) - \frac{d}{d\tau}\partial_{(i}B_{j)}\right] = 0\,.
\end{multline}
The complex ellipticity $\ep$ and the image rotation $\om$ are then obtained from integrating
\be\label{e:ellipom}
\frac{d\ep}{d\tau} = 2 \frac{\tilde{\si}}{\tilde{D}_A^2} \sqrt{4+\ep\ep^\ast}\,, \qquad \frac{d\om}{d\tau} = \tilde{D}_A^{-2} \frac{\ii\left(\ep^\ast \tilde{\si} - \ep \tilde{\si}^\ast\right)}{4 + 2 \sqrt{4+\ep\ep^\ast}}\,.
\ee

The set of coupled ordinary differential equations (\ref{e:photk0}), (\ref{e:photki}), (\ref{e:SachsDA}), (\ref{e:screenpt}), (\ref{e:Sachssi}) and (\ref{e:ellipom}) fully determines our weak-lensing observables once we have fixed the boundary conditions of each ray. At the observer location (assumed to be in the cosmological rest frame\footnote{Once the observables are determined in one frame, they can always be translated to a different frame at the same spacetime location by applying an appropriate Lorentz transformation. Note that this affects the observed redshift (due to the Doppler effect) as well as the observed position and the area distance (due to aberration).}) the boundary conditions are
\be
\tilde{D}_A = 0\,, \qquad \frac{d\tilde{D}_A}{d\tau} = -\er^{\phi+\psi}\,, \qquad \tilde{\si} = \ep = \om = 0\,.
\ee
The boundary condition for $\tilde{k}^0$ is arbitrary and corresponds to some reference frequency that may be used to determine the redshift of the observed source. For convenience we set $\tilde{k}^0 = a \er^{-2\phi-\psi}$ at the observer, in which case the observed redshift of the source simplifies to
\be
1 + z = \left.\left(\sqrt{q^2 \er^{2\phi} + a^2} - n^i q_i \er^\phi\right) \frac{\tilde{k}^0}{a^2} \er^{2\phi+\psi}\right\vert_\mathrm{source}\,,
\ee
where $q_i$ is the canonical peculiar momentum per unit mass of the source. The proper observed area distance $D_A$ is obtained by inverting eq.~(\ref{e:confDA}) at the source location.

As we will explain in more detail in the next subsection, our ray-tracing method determines the boundary condition for $n^i$ for each source through a shooting algorithm. For any given $n^i$ a screen basis at the observer is constructed by choosing $\tilde{e}^i_A$ appropriately.

\subsection{Ray tracing}\label{ss:ray}

The weak-lensing observables $D_A$, $\ep$ and $\om$ are obtained by integrating the coupled ordinary differential equations backwards in time from the observation event. The other endpoint of the integration corresponds to an event on the observer's past light cone. A clear physical interpretation arises, for instance, if that event belongs to the world line of a source (\eg a galaxy). Furthermore, in order to have a notion of observed redshift, one also needs to specify a four-velocity vector at the source location that identifies the source's rest frame. In this work we will consider cold dark matter elements ($N$-body particles) as the source population. This has the advantage that the bias is trivial, and that there are enough sources available even inside large voids. The issue of bias is very interesting and certainly deserves a separate study.

In the weak-lensing regime, for a given observation event and timelike source world line in the same causal patch of the Universe, there exists exactly one past-directed null ray that connects the two. The situation is different for strong lensing, where there can be multiple images, but we assume that the number of sources for which this happens is negligible.
In order to identify that null ray (or one of the possible null rays in the rare case of strong lensing) we use a shooting algorithm similar to the one of \citet{Breton:2018wzk} --- see in particular their Figure 1 for an illustration.

First, for each source we identify the spacetime event at which its world line would cross the past light cone of the observation event in the absence of metric perturbations. In fact, this is how the light-cone data for particles is constructed ``on the fly'' during the simulation run. The background FLRW model also provides us with an initial guess for the boundary condition of $\bn$ at the observer. We then integrate the null geodesic equation in the fully perturbed metric (which is a separate simulation output as explained in the next section) until the ray passes close to the source world line. The source four-velocity allows us to construct a linear segment of the world line, which is sufficient for our purposes. Given the angular distance (estimated as $\bar{D}_A$), the spatial separation of the two geodesics corresponds to a ``deflection angle'' by which the shooting direction $\bn$ must be corrected. The timelike separation corresponds to a ``Shapiro delay'' by which the source has to be moved along its world line in order to meet the null ray. Due to the non-relativistic peculiar velocities the latter correction is minute but we take it into account anyway. With these corrected boundary conditions we integrate the null geodesic equations anew, obtaining a ray that passes significantly closer to the source. The whole process is then iterated a few times until convergence is achieved.

\subsection{Lensing potential}
\label{lens-pot-maps}

We have explained in the last two sections how weak-lensing observables can be constructed non-perturbatively from simulations.
In this section we make contact with the well-known perturbative approach to determine the (linear) convergence $\ka_\mathrm{lin}$ and shear $\gga_1 + \ii \gga_2$.

In the context of perturbation theory, weak lensing is often discussed in terms of the ``lens map'' that maps the direction vectors on the observer's sky to points on a distant ``source plane'' which is in fact a patch of a two-dimensional spacelike sub-manifold. It is important to emphasise the difference between this map and the Jacobi map that maps those direction vectors to points on a \textit{screen} at the source plane. The two maps are only equivalent once we take into account the coordinate transformation between the coordinate basis used to label points on the source plane and the Sachs basis used to label the points on the screen, see Appendix~\ref{a:A}.

With sources at a fixed comoving distance $r$ the points on the source plane can be labelled $r \theta^i_\mathrm{s}$, where $\theta^i_\mathrm{s}$ is a direction vector on the two-sphere. In the weak-lensing regime the lens map provides a one-to-one correspondence between $\theta^i_\mathrm{s}$ and a direction on the observer's sky, $\theta^i_\mathrm{o}$, and we can write
\be
    \theta^i_\mathrm{s} = \theta^i_\mathrm{o} + \alpha^i\,,
\ee
where $\alpha^i$ is called the deflection angle. Its definition is coordinate dependent because $\theta^i_\mathrm{s}$ entirely depends on the chart used at the source plane. The conceptual utility of the deflection angle therefore also hinges on the subtle (often implicit) assumption that the perturbations of the \textit{induced} metric on the source plane can be neglected.\footnote{Probably the most extreme case to illustrate this coordinate dependence are the so-called geodesic light-cone coordinates \citep{Gasperini:2011us}. Here the deflection angle vanishes by construction, and all the information is contained in the chart on the source plane, leading to a highly non-trivial induced metric. The Sachs basis, on the other hand, always provides a basis of vector fields in which the metric is locally Minkowski, and hence the optical parameters describe the proper beam geometry.} In the literature, the Poisson gauge coordinates are often assumed which then gives a unique notion of $\alpha^i$, and from now on we shall follow this convention.

To first order in the metric perturbations $\psi$, $\phi$ and $B_i$ the deflection angle can be computed by integrating eqs.~(\ref{e:nullcondition}) and (\ref{e:photki}),
\begin{multline}
    \alpha^i = \\ \int\limits_0^r \!\frac{dr'}{r} \biggl[
    \left(r-r'\right)\! \left(n^i n^j - \delta^{ij}\right) \partial_j \!\left(\phi + \psi + n^k B_k\right) - \delta^{ij} B_j\biggr]\,.
\end{multline}
If we neglect the frame-dragging effect and furthermore insert the unperturbed trajectory in the integrand, the so-called Born approximation, which means $x^i(r') \simeq r' \theta^i_\mathrm{o}$, we obtain
\be
\alpha_a = -\nabla_a \int\limits_0^r dr' \frac{r-r'}{rr'} \left(\phi+\psi\right)\,,
\ee
where the index $a$ now refers to coordinates on the two-sphere (corresponding to $\theta^i_\mathrm{o}$ and hereafter denoted as $\bth$). The fact that
$\alpha_a$ can be written as a gradient field under this approximation motivates the definition of the
lensing potential \citep{Lewis:2006}
\be\label{e:lensPot}
\Psi(\bth,z) \equiv-\int\limits_{0}^{r(z)} dr' \frac{r(z)-r'}{r(z) r'} \left(\phi + \psi\right)\,.
\ee

According to eq.~(\ref{e:screenpt}) the Sachs basis does not rotate (with respect to the coordinate basis) at leading order in the absence of frame dragging. Therefore, the lens map and the Jacobi map are both approximated by
\be
\AAA = \id + \left(\nabla_a \nabla_b \Psi\right)\,,
\ee
and hence
\begin{align}
\label{e:kappa_per}
\ka_\mathrm{lin} &= -\frac{1}{2}\De\Psi\,,\\
\gga_1+i\gga_2 &= -\frac{1}{2}\left(\nabla_1\nabla_1 - \nabla_2\nabla_2\right) \Psi -\ii\nabla_1\nabla_2\Psi\,.\qquad
\end{align}
The complex shear is a spin-2 quantity which we can decompose into its ``electric'' component, $\gga_E$, which has positive parity and its ``magnetic'' component, $\gga_B$, which has negative parity, see \citet{Bernardeau:2009bm}.
It is easy to verify that in the above case which is relevant within linear perturbation theory (in the absence of frame dragging and gravitational waves) $\gga_B\equiv 0$.

Once we include frame dragging or go beyond leading order and take into account \eg post-Born corrections, the deflection angle is in general no longer a gradient map. We can decompose it
into a gradient and a curl component,
\be \label{e:gradmap}
\al_a = \nabla_a\Psi + \varepsilon_a^b\nabla_b\Om = (\bnabla\Psi +\bnabla\wedge\Om)_a \,.
\ee
Here $\varepsilon_a^b$ is the totally anti-symmetric tensor in two dimensions. Note that in two dimensions the curl of a vector is a scalar while the curl of a (pseudo-)scalar as defined above is a vector.
Evidently, if $\Om \neq 0$ the lens map has an antisymmetric part, but we already know that the image rotation $\om$ vanishes at first order in any spacetime. Indeed, if we look at the first-order calculation including $B_i$ we have to take into account the fact that the Sachs basis rotates with respect to the coordinate basis. It is easy to verify \citep{Dai:2013nda,DiDio:2019rfy} that this rotation exactly cancels the one from the lens map at leading order, so that the Jacobi map is again symmetric. This also means that the polarisation (which is parallel-transported in the same way as the Sachs basis) rotates in the Poisson gauge, which has interesting consequences for CMB observables.

It should, however, be evident that the rotation of the Sachs basis does not change the shear signal at all for a statistically isotropic and uncorrelated source population (what makes the effect relevant to the CMB is the fact that the sources have intrinsic correlations).
In linear theory \citet{Hirata:2003ka} showed that in the ``flat-sky approximation''
\begin{align}
C_\ell^{\gga_E} &= \frac{\ell^4}{4}C_\ell^\Psi = C_\ell^{\ka}\,, \\
C_\ell^{\gga_B} &= \frac{\ell^4}{4}C_\ell^\Om \,,
\end{align}
where the $C^X_\ell$ are angular power spectra as defined in Sec.~\ref{Cell} below. \citet{Hirata:2003ka} also find that the rotation of the lens map (\ie its antisymmetric part) has the same power spectrum as $\gga_B$, but they do not take into account the fact that the Sachs basis rotates. The rotation of the lens map is not directly observable (because it is gauge-dependent), as opposed to the image rotation in the Sachs basis which can be observed, \eg for a polarised source where polarisation and ellipticity are aligned. We note that frame dragging does produce an observable B-mode shear at first order, but no image rotation (in the sense that $\omega$ vanishes). However, in standard cosmological perturbation theory the frame-dragging effect itself only appears at second order \citep{Lu:2008ju}, and hence its contribution is typically of the same order as post-Born corrections and other second-order contributions that do allow for image rotation to occur.

In Newtonian simulations the lensing potential is often constructed from mass maps [see \citet{Hilbert:2019vca} for a comparison of different state-of-the-art numerical methods]. Here one uses that in the Newtonian sub-horizon limit the two Bardeen potentials are equal and related to the matter over-density $\delta_m$ through the Poisson equation
\be
\label{e:Poisson}
\partial^2 \phi = \partial^2 \psi = \frac{3 H_0^2 \Omega_m}{2} \left(1+z\right) \delta_m\,.
\ee
Together with equations (\ref{e:lensPot}) and (\ref{e:kappa_per}) one then finds
\be
\ka_\mathrm{lin} = \int\limits_{0}^{r} dr' r' \frac{r-r'}{r} \frac{3 H_0^2 \Omega_m}{2} \left[1+z(r')\right] \delta_m - \left.\frac{\phi+\psi}{2}\right\vert_0^r\,,
\ee
where the boundary term is usually dropped because it is $\sim 10^{-5}$ and therefore much smaller than the typical value of $\ka$. We note, however, that this approximation would induce spurious effects if one cross-correlates the lensing signal with the gravitational redshift of the sources.

It is worth pointing out that this type of lensing analysis is only applicable if the gravitational fields are indeed generated by matter according to eq.~(\ref{e:Poisson}), and therefore precludes modifications of gravity or the possibility that in general there can be sources to metric perturbations other than nonrelativistic matter. Our geometric approach, working directly with the metric perturbations, is therefore more robust and flexible.

Furthermore, it has recently been demonstrated \citep{Fidler:2017pnb,Adamek:2019aad} that the weak-lensing analyses of Newtonian $N$-body simulations would have to include a gauge correction that is commonly neglected. This is due to the fact that the coordinate system of the Newtonian simulations does not coincide with the Poisson gauge for which the lensing calculations have been developed. The correction appears at large angular scales and can be incorporated into a modified lensing potential. This is not an issue in our relativistic simulations as we use Poisson-gauge coordinates consistently throughout.

We want to study how well the first-order lensing potential (neglecting frame dragging) characterises the full weak-lensing signal. To this end we construct maps of the lensing potential at fixed comoving distance (corresponding to fixed redshift in the background model) by numerically integrating eq.~(\ref{e:lensPot}). Due to the Born approximation this can be done very efficiently directly in pixel space, as the gravitational potentials $\phi$ and $\psi$ are already conveniently pixelised (this is explained in more detail in the next section). Once the lensing potential $\Psi$ is computed, we generate a map of $\ka_\mathrm{lin}$ by solving eq.~(\ref{e:kappa_per}). This functionality is readily available in the public release of \textit{gevolution} version 1.2.\footnote{\url{https://github.com/gevolution-code/gevolution-1.2}} Appendix \ref{a:validation} presents a direct validation against other weak-lensing codes within the Newtonian approximation.

\section{Simulation}

Our numerical results, presented in the next section, are based on a large $N$-body simulation using the relativistic code \textit{gevolution} \citep{Adamek:2015eda}. The simulation has $7680^3$ (almost half a trillion) mass elements in a cosmological volume of $(2.4\, \mathrm{Gpc}/h)^3$, which gives a mass resolution of $2.6 \times 10^9\, M_\odot / h$. We use a standard cosmology with $\Omega_m = 0.312$, $\Omega_b = 0.048$, $h = 0.67556$, and massless neutrinos with $N_\mathrm{eff} = 3.046$. We generate the linear transfer functions of baryons and cold dark matter at initial redshift $z_\mathrm{ini} = 127$ with \textsc{class} \citep{Blas:2011rf}, and choose a primordial amplitude of scalar perturbations $A_s = 2.215 \times 10^{-9}$ at the pivot scale  $k_*=0.05\, \mathrm{Mpc}^{-1}$ and spectral index $n_s = 0.9619$. This corresponds to $\si_8 = 0.8488$.

We choose an observer located in the corner of the box and consider a pencil beam on the past light cone, with a $450$ sq.\ deg.\ field-of-view centered around $(l = 33.4^\circ, b = 48.6^\circ)$ in ``galactic coordinates'' aligned with the principal axes. In a small region up to a comoving distance of $275\, \mathrm{Mpc}/h$ we retain the past light cone on the full sky, in order to have more data around the apex of the pencil beam. This specific geometry was chosen such that the conical footprint of the mock survey does not contain any replications up to redshift $z \simeq 3.25$. In other words, the pencil beam is allowed to pass through the periodic domain more than once, but without any self-intersection. Finite-volume effects can, however, not be avoided entirely, \eg the fact that no modes larger than the fundamental mode can contribute to the perturbations. We note that the pencil beam makes its first box crossing at redshift $z \simeq 1.65$ which means that observables below that redshift  are not affected. Even at higher redshift we expect the effect of fake correlations between distant portions of the light cone to be insignificant for our analysis, given that our mock survey uses only about $32\%$ of the full simulation volume and that correlations decay rapidly with increasing distance.

The data on the light cone are recorded as follows. The comoving positions and the peculiar momenta of $N$-body particles are recorded at the time when their world lines would intersect the past light cone if the metric was unperturbed. Up to a small Shapiro delay this intersection will be very close (in terms of four-dimensional space-time distance) to the intersection with the true light cone of the perturbed metric. The linear segment of the world line that can be constructed from the particle phase space coordinates is therefore sufficient for locating that latter intersection point in the later analysis.

The true past light cone can only be determined from a complete knowledge of the metric, which is only available at the end of the simulation. The metric perturbations are initially sampled on a Cartesian mesh with a comoving spatial resolution of $312.5\, \mathrm{kpc}/h$. However, for the purpose of constructing a light cone it is more convenient --- and computationally much more efficient --- to work in spherical coordinates. We therefore define a second coordinate mesh, consisting of concentric spherical shells with the observer at the origin. The shells are separated radially by our base resolution of $312.5\, \mathrm{kpc}/h$, and are each pixelised using the \textsc{HEALPix} framework \citep{Gorski:2004by}. The number of pixels is chosen for each shell separately with the requirement that the area of each pixel is always less than $(312.5\, \mathrm{kpc}/h)^2$. Since the number of pixels can only be changed discontinuously this means that our \textsc{HEALPix}-mesh is generally denser than the Cartesian mesh.

For each time step of the simulation we first identify the corresponding comoving distance interval on the past light cone in the unperturbed spacetime. We then record the perturbed metric on all the shells within that distance interval by ``triangular-shaped particle'' interpolation from the Cartesian mesh onto the pixel locations. The same is done for the preceding as well as the following time step, such that we effectively retain a finite portion of the four-dimensional spacetime around the light cone. This allows us to reconstruct the true light cone without any approximations, by tracing null-geodesics backwards in time from the observer.

\section{Results}\label{s:res}

In this section we discuss the main results of our work. 
The outline of the procedure is as follows:
(1) we perform ray-tracing over a subset of the particles in our simulation, (2) we compute 
convergence, shear and rotation angle for each source in our catalogue, (3) we project 
these fields onto a pixelised sky map, (4) we
compute the angular power spectra for the maps
of the fields for different redshifts, (5) we compute the one-point probability distribution function (PDF) for the values of the fields.

For our analysis, we run the ray tracer
over $\sim 135$ million particles, randomly selected from the full $N$-body ensemble. 
The reason to use a particle catalogue rather
than a halo catalogue is twofold: first, the $N$-body particles are an unbiased tracer of the dark matter and second, the larger sample size offers better 
statistics compared to a halo-based study. 

\begin{figure}
\centering
\includegraphics[width=\columnwidth]{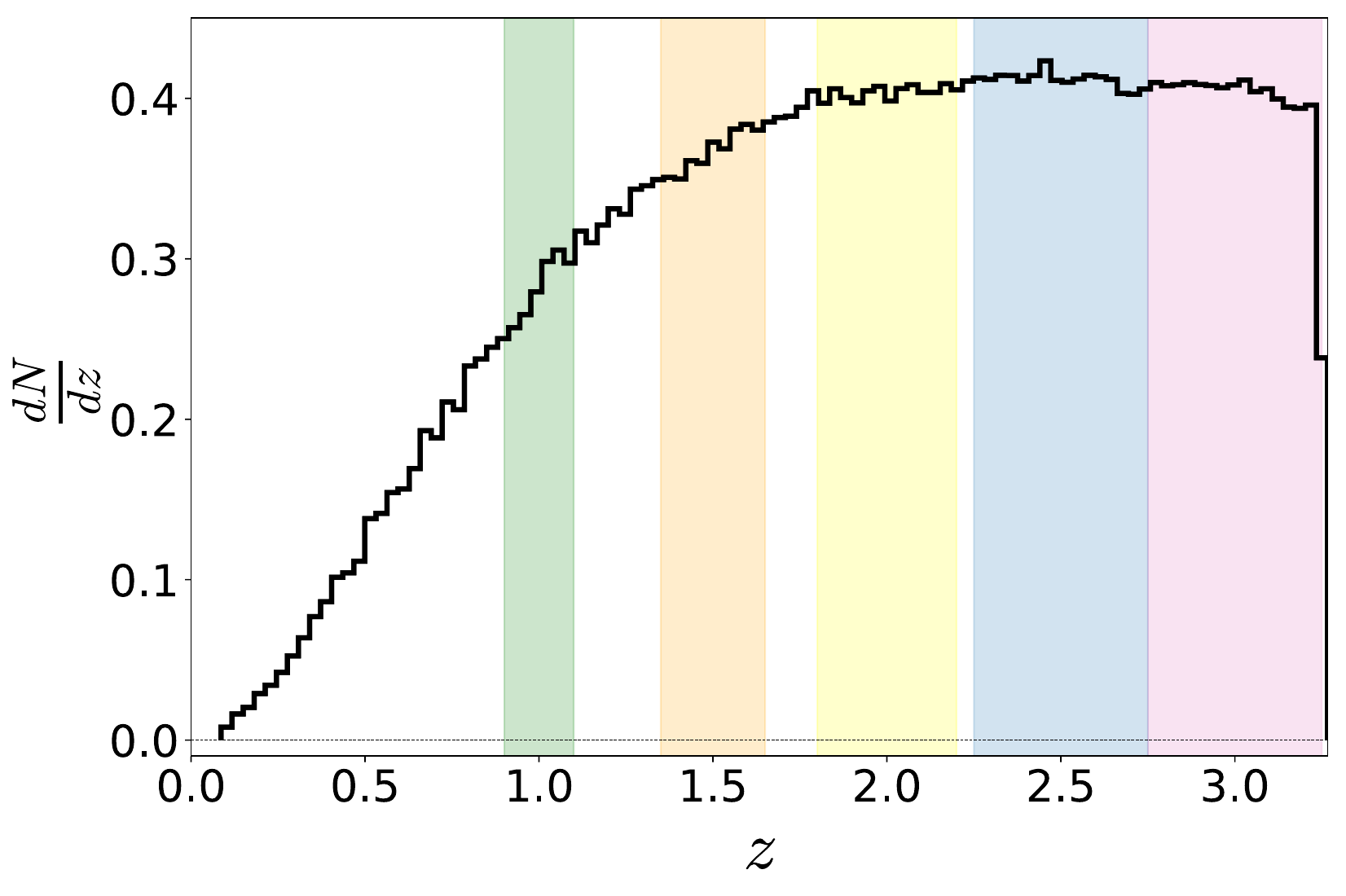}
\vspace{-10pt}
\caption{Redshift distribution of the sources. The redshift bins used in our analysis (see Table \ref{table:1}) are indicated as shaded colour bars.}
\label{fig:catalogue}
\end{figure}

\begin{figure*}
\centering
\subfigure[Convergence $\ka$]{\includegraphics[width=0.46\textwidth]{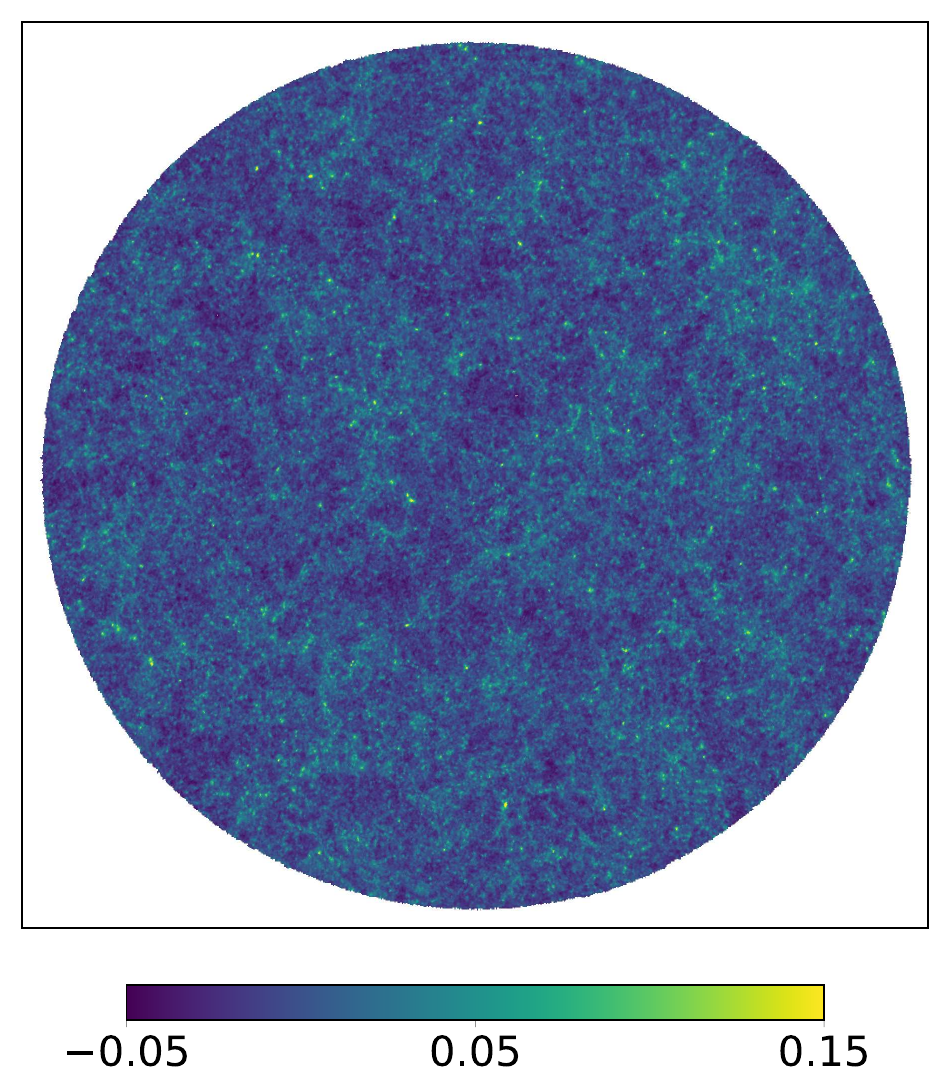}}
\subfigure[Elliplicity $|\ep|$]{\includegraphics[width=0.46\textwidth]{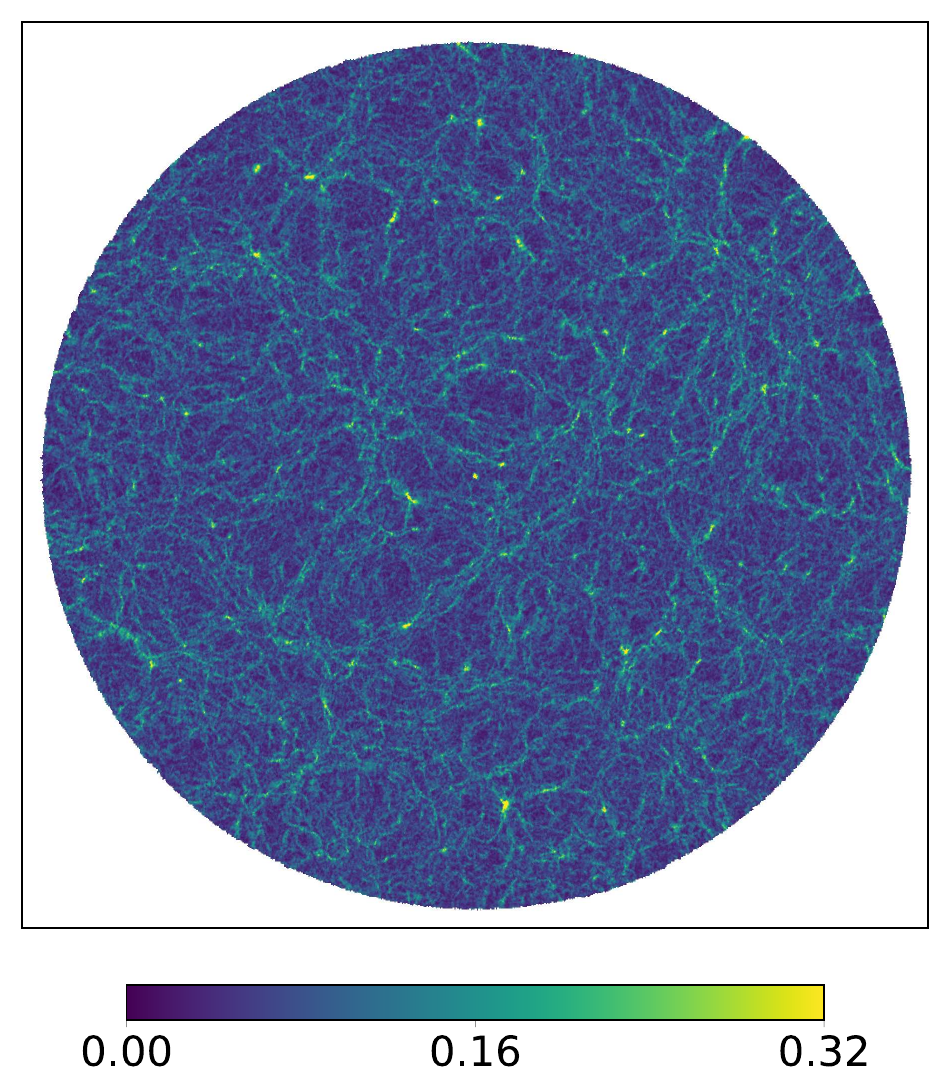}}\\
\subfigure[Magnification $\mu=\frac{1}{(1-\ka)^2}$]{\includegraphics[width=0.46\textwidth]{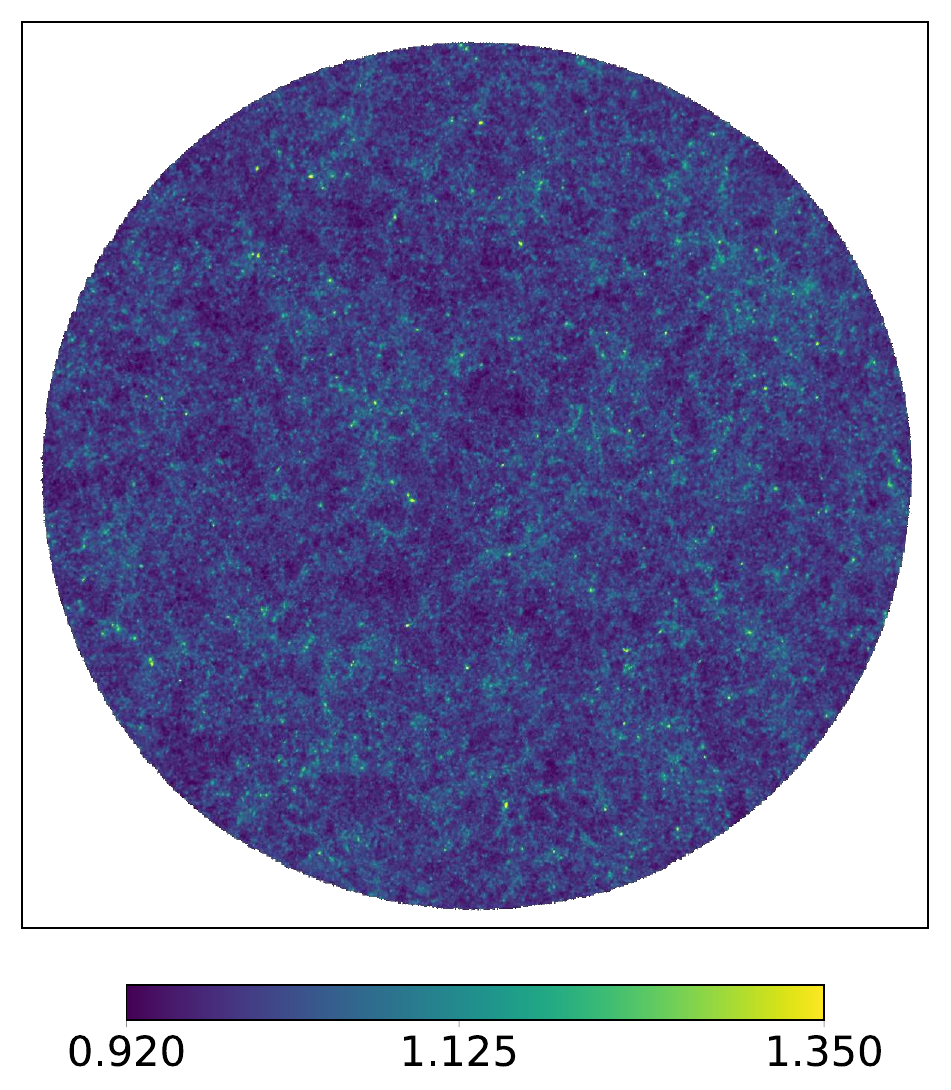}}
\subfigure[Rotation angle $\omega$]{\includegraphics[width=0.46\textwidth]{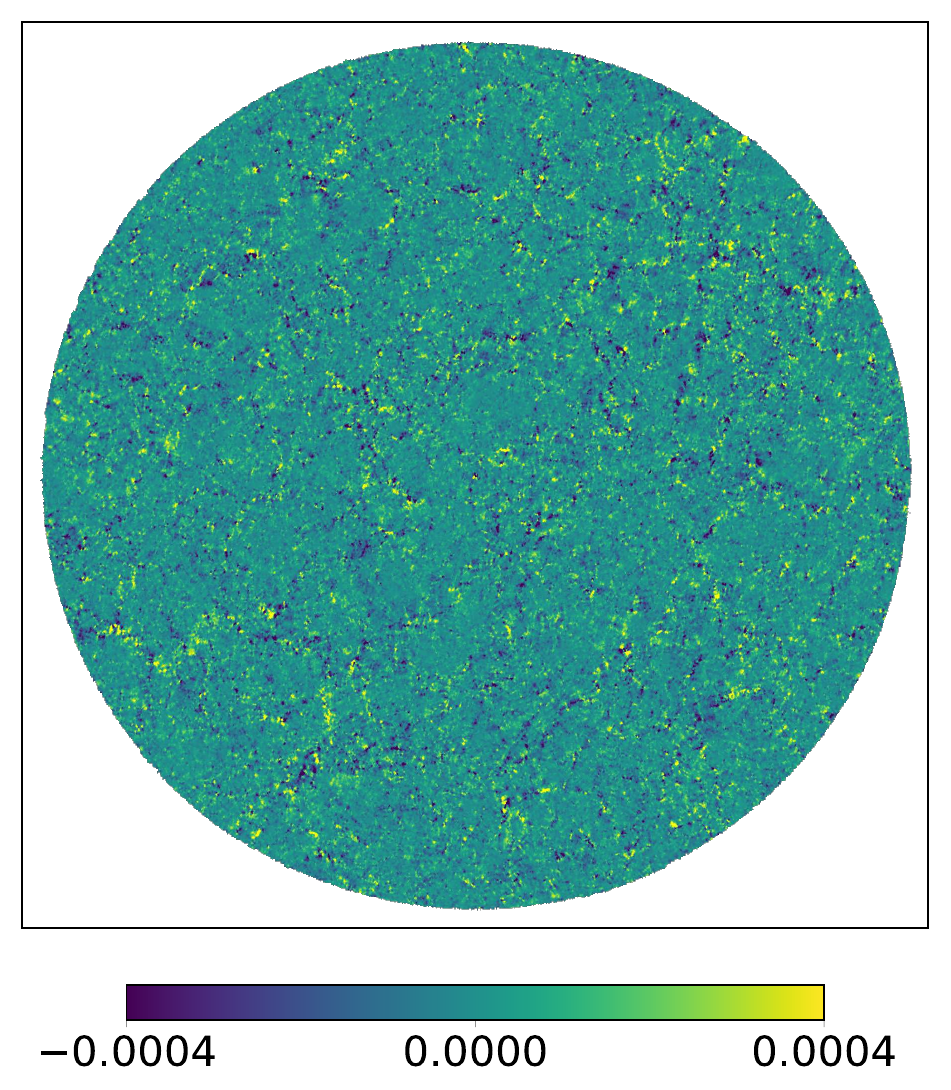}}
\caption{Maps for the weak-lensing fields at $z = 1.5$ (gnomonic projection).
The patch of the sky displayed in the
figure subtends about 24 degrees, or a solid angle of approximately 450 square degrees.
}
\label{fig:maps}
\end{figure*}

The normalised redshift distribution of the
particle catalogue is show in Fig.~\ref{fig:catalogue}.
The particle catalogue covers a redshift range
between $z=0$ and $z \simeq 3.25$.
At low-redshift, up to $z \simeq 0.09$,  the particles are distributed over the whole sky,
while for $0.09 \lesssim z \lesssim 3.25$ the particles are distributed over a sky fraction $f_\mathrm{sky} = 0.01$.
We focus our analysis on five 
redshift bins (see Table~\ref{table:1}).
\begin{table}
\begin{center}
\begin{tabular}{|c c c| } 
 \hline
 $z$ & $\delta z$ & $N_\text{sources}$\\ 
 \hline
 1   & 0.1 & 7531359\\ 
 1.5   & 0.15 & 14772881\\ 
 2   & 0.2 & 21630114\\ 
 2.5   & 0.25 & 27581488\\ 
 3   & 0.25 & 27104790\\ 
 \hline
\end{tabular}
\caption{The mean redshifts and half-widths of 
the  bins considered in our analysis. We also quote the number of observed sources $N_\text{sources}$ in each redshift bin.\label{table:1}}
\end{center}
\end{table}
These bins are also outlined in Fig.~\ref{fig:catalogue} as vertical colourbars.

The remainder of this section is organised as
follows. 
In Sec.\ \ref{maps} we describe the method to
obtain field maps from the optical parameters of the source population, in Sec.\ \ref{Cell} and Sec.\ \ref{pdf} we discuss the weak-lensing angular power spectra and PDFs, respectively, as well as their redshift dependence.
In Sec.\ \ref{born} we compare the non-linear convergence
obtained from the ray tracer to the linear convergence computed within the Born approximation (\ie along the unperturbed light path), from the lensing potential map.

\subsection{Weak-lensing maps}
\label{maps}

The ray tracer numerically computes the area distance, the complex ellipticity and the rotation angle of the image, at the observed redshift
and angular position in the sky, for all the dark matter particles in the catalogue. 
In our procedure, we work only with
observable coordinates, the observed sky position and redshift. Once these observables have been computed, neither the comoving position nor the peculiar velocity of the $N$-body particles is used in the analysis, as these would be specific to the Poisson gauge and therefore have no invariant meaning.

Convergence and shear, at the observed position
of each particle, are estimated from eq.~\eqref{conv-shear}. The convergence depends only on the area distance, while the shear coincides with the ellipticity, up to a  factor $4$. The magnification is a function of the convergence alone and it is computed from eq. \eqref{magn}.

Convergence, ellipticity (respectively its absolute value), magnification and rotation
angle are projected onto a
\textsc{HEALPix} map in the following way: at fixed redshift $z$, we select all the particles inside 
the redshift bin of thickness $\delta z$ and the value of the field in each pixel is computed as the average value from all the particles inside the pixel volume.

In this section, we adopt a map resolution of $N_\text{side} = 2048$ which corresponds to an angular resolution of $\sim 1.71$ arcmin.
The number of pixels over the whole sky is $N_\text{pixel} = 12 \times N_\text{side}^2$ which means that our survey area provides approximately $0.5$ megapixels. 
In Appendix \ref{res-test} we discuss the impact of angular resolution on our results. 

In Fig.~\ref{fig:maps} we show the maps for the convergence, the  amplitude of the ellipticity, $|\ep|$, the magnification and the rotation angle at mean redshift $z=1.5$. 
The maps highlight that convergence and magnification show very similar patterns. 
Ellipticity and rotation angle show a significantly different small-scale behaviour. However,
the regions in the map with higher ellipticity and $\omega$ (in absolute value)
correspond to the regions in the maps with
larger convergence and magnification. Despite the fact that ellipticity and convergence have nearly identical spectra, as expected from the linear analysis,  their maps look quite different with elongated ``filaments'' in the ellipticity map which are not present in the convergence map. These patterns are certainly related to similar ones that have been found for the deflection angle \citep{Carbone:2007yy,Watson:2013cxa}.

\subsection{Angular power spectra}
\label{Cell}
Angular power spectra for our observables
can be obtained through spherical harmonic
decomposition.
Convergence and rotation, being 
scalar and pseudo scalar quantities, can be expanded in 
spherical harmonics
\begin{align}
\ka(\bth, z) &= \sum_{\ell m}
\ka_{\ell m} (z) Y_{\ell m}(\bth)\,,\\
\omega(\bth, z) &= \sum_{\ell m}
\omega_{\ell m} (z) Y_{\ell m}(\bth)\,,
\end{align}
where $\ka_{\ell m}$ and
$\omega_{\ell m}$ are the coefficients of the spherical harmonic decomposition.

In a similar way, the ellipticity $\ep$ which is a spin-2 object, can be expanded in $\pm 2$ spin weighted spherical harmonics \citep{Chon:2003gx}
\begin{align}
\ep(\bth, z) &= 
\sum_{\ell m} \left( \ep^E_{\ell m}(z) +i
 \ep^B_{\ell m}(z)\right)\, {}_{2} Y_{\ell m}(\bth)\,, \\
 \ep^*(\bth, z) &= 
\sum_{\ell m} \left( \ep^E_{\ell m}(z) -i
 \ep^B_{\ell m}(z)\right)\, {}_{- 2} Y_{\ell m}(\bth)\,,
\end{align}
where $\ep^E$ and $\ep^B$ denotes
the E and B modes of the ellipticity, respectively.

The angular power spectra, at fixed
redshift\footnote{We omit the redshift dependence for brevity.}, are defined as follows
\begin{align}
\mean{\ka_{\ell m}\ka_{\ell' m'}}
&= \delta_{\ell \ell'} \delta_{m m'}
C_\ell^{\ka}\,, \\
\mean{\omega_{\ell m} \omega_{\ell' m'}}
&= \delta_{\ell \ell'} \delta_{m m'}
C_\ell^{\omega}\,,
\end{align}
\begin{align}
\mean{\ep^E_{\ell m} \ep^E_{\ell' m'}}
&= \delta_{\ell \ell'} \delta_{m m'}
C_\ell^{\ep_E}\,, \\
\mean{\ep^B_{\ell m} \ep^B_{\ell' m'}}
&= \delta_{\ell \ell'} \delta_{m m'}
C_\ell^{\ep_B}\,.
\end{align}

The maps we obtain from our simulations
and ray-tracing cover about $1\%$ of the sky. Therefore, in the spherical harmonic
decomposition outlined above, masking effects
are crucial and need to be corrected for.

In fact, the standard \textsc{HEALPix} routine for the angular power spectrum estimation  - \textsc{anafast} - sets to zero the masked pixels before the spherical harmonic decomposition and would introduce a large bias in our analysis.

\begin{figure}
\centering
\includegraphics[width=\columnwidth]{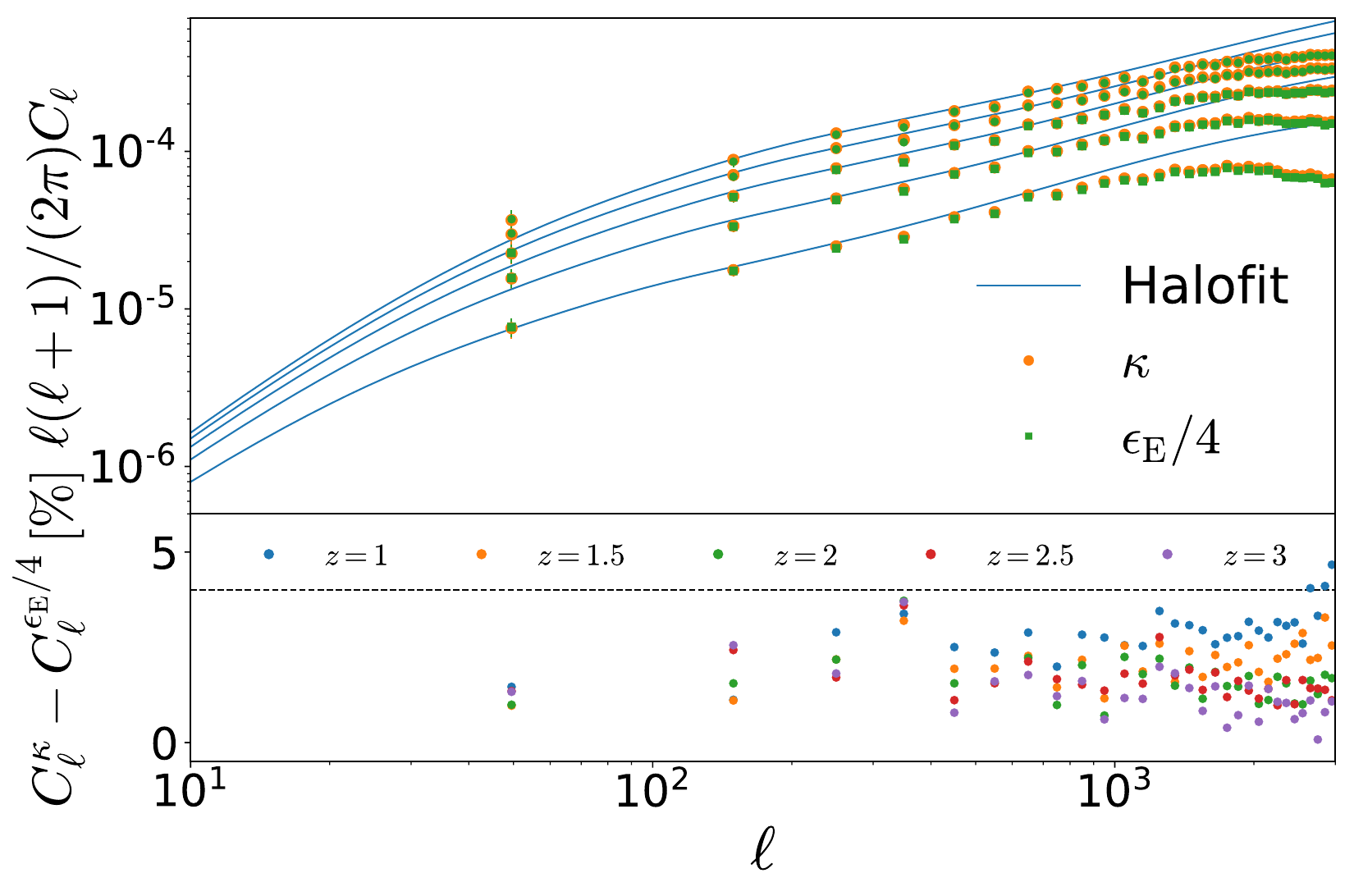}
\vspace{-10pt}
\caption{Top panel: Angular power spectra of $\ka$ and $\ep_E/4$ at redshifts $z=$ 1, 1.5, 2, 2.5 and 3 from bottom to top. Bottom panel: Relative difference between the $\ka$ and $\ep_E/4$ power spectra in \%. We plot the spectrum of $\ep_E/4$ and not $\ep_E$ since within linear perturbation theory this spectrum agrees with the $\ka$-spectrum.
The dashed line corresponds to 4\% relative difference. 
}
\label{fig:kappagamma}
\end{figure}

In order to correct for the masking effects,
we estimate the angular power spectra for our maps with the code \textsc{PolSpice}, which
implements an unbiased estimator for the power spectra based on the estimation of weighted 
correlation functions. The details on this estimator can be found in  \citet{Szapudi:2000xj, Chon:2003gx}. 
The estimated power spectra are binned linearly in $\ell$ with bin size $\Delta \ell = 100$, which
corresponds to $1/f_\text{sky}$. In this way, we reduce the cosmic variance errors and spurious oscillations in the spectra.

In Fig. \ref{fig:kappagamma} we show the estimated angular power spectra for 
the convergence and the ellipticity E-modes. In the perturbative approach,
the two power spectra are
related through\footnote{This is derived, \eg in~\cite{Montanari:2015rga}.}
\be
C^{\ep_E/4}_\ell =
\frac{(\ell+2)(\ell-1)}{\ell(\ell+1)} C^{\kappa}_\ell\,.
\ee
Therefore, we expect $C^{\ep_E/4}_\ell \approx C^{\kappa}_\ell$
on large multipoles. 
On the top panel, we compare the numerical convergence and ellipticity spectra with the expected power spectrum 
estimated from {\sc class}, where non-linearities are taken into account through
the \textsc{Halofit} corrections to the power spectrum \citep{Takahashi:2012em}. 
Note that at $\ell \gtrsim 1000$ the spectra estimated from our simulations show a significant loss of power. 
In  Appendix \ref{res-test} we investigate the cause of this suppression and we find that it is mainly caused by the finite resolution of the simulation mesh. 
On the bottom panel, we show the relative difference between the convergence and 
ellipticity E-modes (rescaled by a factor 4). The relative difference is smaller than $5\%$ at all scales and for all the redshift bins.

As discussed in Sec.~\ref{lens-pot-maps}, 
at linear order in perturbation theory shear B-modes and the rotation angle $\omega$ are identically zero if frame dragging is neglected (it is usually considered as a second-order effect).
However, post-Born corrections do source both these quantities and the curl-mode of the ellipticity also receives contribution from the vector perturbations in the metric. 

\begin{figure}
\centering
\includegraphics[width=\columnwidth]{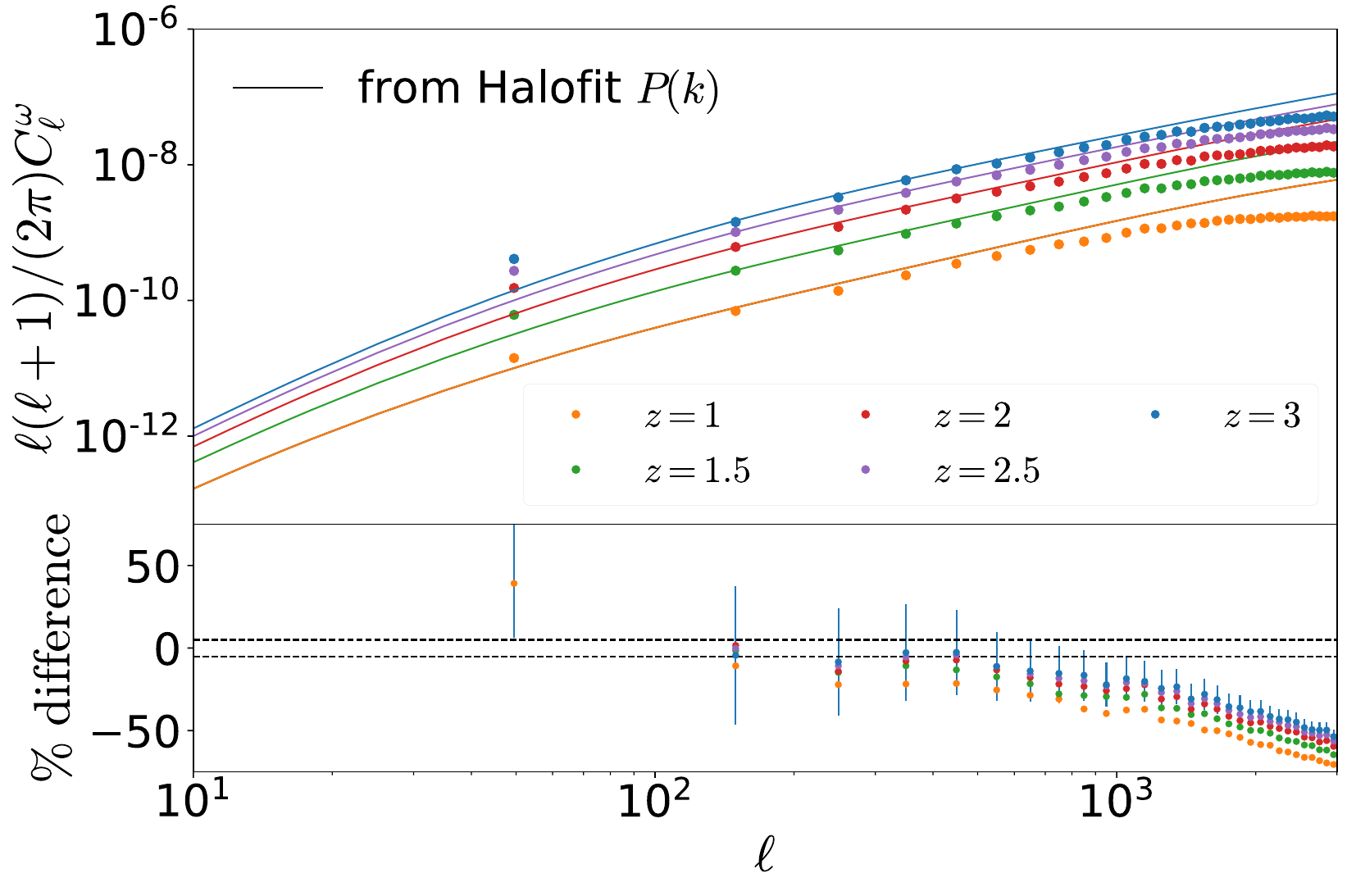}
\vspace{-10pt}
\caption{Top panel: Angular power spectrum for the rotation $\omega$. The circular markers represent the results from our simulations, while continuous lines are the post-Born predictions computed with \textsc{CAMB}. Bottom panel: Relative difference between the power spectrum estimated from the simulation and the theoretical prediction from second-order perturbation theory (using the \textsc{Halofit} matter power spectrum). 
The horizontal dashed lines represent 
$\pm 5\%$ relative difference.
Different colors denote different redshifts. 
}
\label{fig:omega-cl}
\end{figure}

Since the rotation $\omega$ is a pseudo-scalar,
extracting its angular power spectrum from a masked map is not particularly problematic.
In Fig. \ref{fig:omega-cl} we show the angular 
power spectrum for the rotation, at different redshifts. Data points refer to the results from
our simulations, while continuous lines are the
theoretical prediction from second-order post-Born lensing calculations \citep{Pratten:2016dsm, Marozzi:2016qxl}. 
The theoretical predictions are computed using the post-Born lensing module implemented in the
Code for Anisotropies in the Microwave Background (\textsc{CAMB})\footnote{\url{https://github.com/cmbant/CAMB}}. Similarly to the convergence and ellipticity spectra, non-linearities are modelled with the \textsc{Halofit} prescription \citep{Takahashi:2012em}. 

On large scales we find good agreement between the result of our simulations and the post-Born 
prediction. On small scales, we notice that 
$\omega$ suffers  a power-loss due to finite grid resolution effects, similar to what we observe for the convergence and  ellipticity spectra. 

The rotation power spectrum has been computed
in previous works, based on Newtonian simulations, using a multiple-lens-plane approximation \citep{Becker:2012qe, Takahashi:2017hjr, Fabbian:2019tik}.
Our results qualitatively agree with previous
results in the literature, where a few-percent agreement was found between the spectra
extracted from Newtonian simulations and 
the second-order post-Born prediction.

At linear order (and neglecting frame dragging), the ellipticity B-modes and the rotation $\omega$ spectra 
both vanish.
Both are generated only at second order and we therefore expect them to be of comparable amplitude (yet not identical -- note in particular that the ellipticity receives a contribution that is linear in the frame-dragging potential, while the rotation does not, see also App.~\ref{a:A}).
However, we are not able to extract the $\ep_B$ component above the noise from our numerical results for the ellipticity. Considering the four orders of magnitude and more difference in amplitude, we believe that this is due to masking and pixelisation effects that inevitably lead to some leakage between $\ep_E$ and $\ep_B$.

\subsection{Probability distribution functions}
\label{pdf}
In this section, we show the results for the one-point Probability Distribution Function (PDF) for the convergence, the magnification, the ellipticity and the rotation.
Our results are complementary to
previous work on this topic [see \eg \citet{Takahashi:2012em}, where a detailed study of the lensing PDFs is presented 
for high-resolution and small-volume simulations]. 

The PDFs discussed is this section have been computed
in `pixel-space', \ie the fields have been estimated on a \textsc{HEALPix} map, as described in Sec.\ \ref{maps}, and we computed the distribution of the field values in the pixels. 

We considered five redshift bins with
mean redshifts in the range $1 \leq z \leq 3$ as detailed in Table~\ref{table:1} and 
a map resolution $N_\text{side} = 2048$.

\begin{figure}
\centering
\includegraphics[width=\columnwidth]{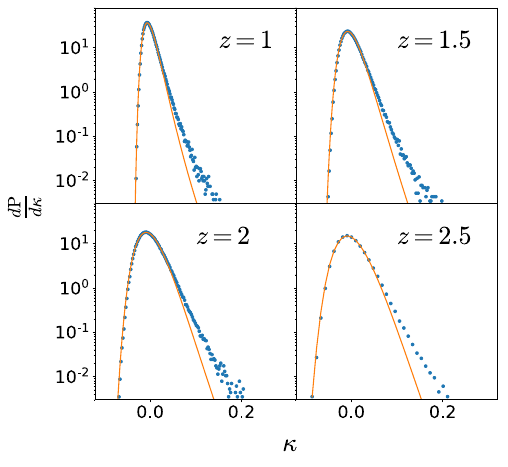}
\vspace{-10pt}
\caption{PDF for the estimated convergence $\ka$ for different redshifts. The lines present the log-normal fit. The fit clearly underestimates the number of high values of $\ka$.}
\label{fig:pdf-kappa-fit}
\end{figure}

In Fig.\ \ref{fig:pdf-kappa-fit} we show the 
normalised PDF for $\ka$ at different redshifts.
The shape of the $\ka$-PDF qualitatively agrees with previous results in the literature \citep{Takahashi:2012em}: the distribution
broadens at high $z$ and the peak position 
 shifts toward smaller values of $\ka$.  
The distribution is strongly skewed towards 
positive values of $\ka$. This is due to the fact that $\ka$ cannot be more negative than its value for an empty beam, $\ka_{\text{empty}}$ given in eq.~\eqref{e:kempty} below. Since modelling the
non-Gaussian shape is crucial in order to
avoid systematic bias in distance--redshift
relation measurements, we have studied the shape of the distribution and its redshift dependence.

\begin{table}
\begin{center}
\begin{tabular}{|c c c c| } 
 \hline
 $z$ & $\sigma_\ka$ & $A_\ka$ & $\kappa_\text{empty}$\\ 
 \hline
 1   & 0.375 & 3.373 & -0.0658 \\ 
 1.5   & 0.298 & 3.587 & -0.129\\ 
 2   & 0.250 & 3.706 & -0.198  \\ 
 2.5   & 0.217& 3.748 & -0.270\\ 
 3   & 0.192 & 3.551 & -0.342 \\ 
 \hline
\end{tabular}
\caption{Parameters of the modified log-normal
 model, fitted to the convergence PDF, as a function of redshift.\label{table:2}}
\end{center}
\end{table}

We consider the modified lognormal distribution, first proposed in  \cite{Das:2005yb} and further validated 
against simulations in \cite{Takahashi:2012em}:
\begin{multline}
\frac{dP}{d\ka} = \frac{N_{\ka}}{\ka + \kappa_\text{empty}} \exp \Biggl[
-\frac{1}{2\sigma^2_\ka} \left\{\ln\left(
1 + \frac{\ka}{\kappa_\text{empty}}\right)
+ \frac{\sigma_\ka^2}{2}
\right\}^2 \Biggr. \\
 \Biggl. \times \left\{
1 + \frac{A_\ka}{ 1 + \ka/|\kappa_\text{empty}|}
\right\} \Biggr],   \label{e:pka}
\end{multline}
where the convergence of the empty beam, $\ka_\text{empty}$ is
\begin{equation}\label{e:kempty}
\ka_\text{empty}(z) = -\frac{3}{2} H_0^2 \Om_m \int^{z}_0 \frac{dz'}{H(z')} (1+z')
\frac{r(z') r(z, z')}{r(z)}\,.
\end{equation}
Here $r(z_1, z_2)$ is the comoving distance 
from $z_2$ to $z_1$ and $r(z) = r(0, z)$. 

The fitted model parameters
are the normalisation $N_\ka$, the parameter $\sigma_\ka$, which determines both the 
width of the distribution of $x=\ka/|\kappa_\text{empty}|$ and the peak position, and the modification to the lognormal distribution $A_\ka$. 

The values of the fitted parameters are reported in Table \ref{table:2}. We omit the value 
of the normalisation. 
Note that the modification to the log-normal 
distribution does not depend strongly on redshift, while $\si_\ka$ decreases at high $z$.
This tells us that the width of the distribution as a function
of $x=\ka/|\ka_\text{empty}|$ is decreasing towards 
higher redshift. However, $\ka_\text{empty}$ is 
is larger at high $z$ so that the width of the
$\kappa$-distribution is increasing (see Fig.\ \ref{fig:pdf-kappa-fit}).

In Fig.\ \ref{fig:pdf-kappa-fit} we show the result of our fits at $z = 1, 1.5, 2, 2.5$.
The modified log-normal distribution 
is able to reproduce accurately the shape of
the convergence PDF near the position of the 
peak and for small convergence values, while 
the fitting function underestimates the PDF
for large values of $\kappa$. The steep decent of the distribution for $\ka\ra \ka_\text{empty}$ is very well captured by \eqref{e:pka}. These results
agree with the analysis in \cite{Takahashi:2012em}.

\begin{figure}
\centering
\includegraphics[width=\columnwidth]{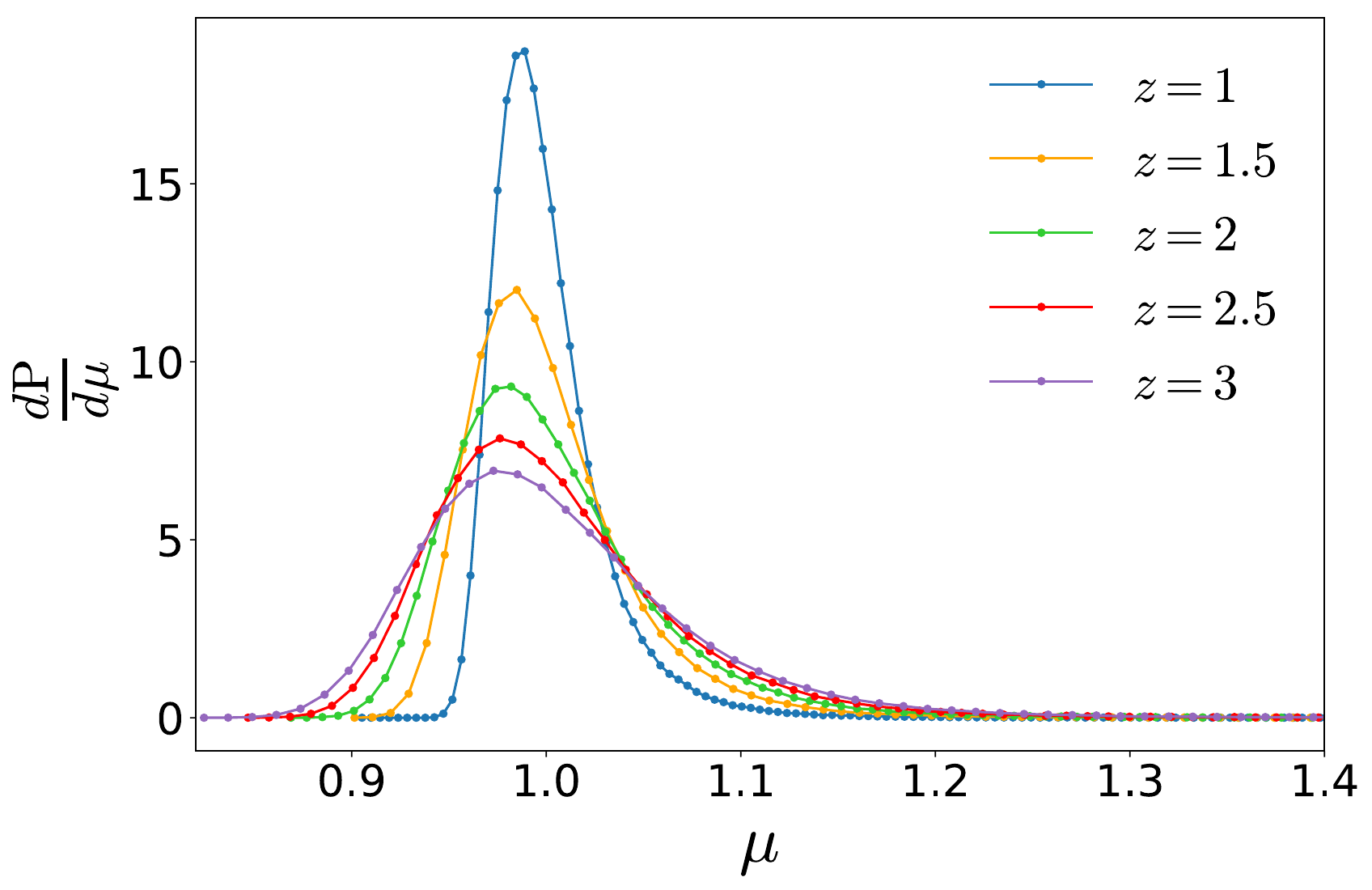}
\vspace{-10pt}
\caption{PDF for the magnification $\mu$ in different redshift bins.}
\label{fig:pdf-mu}
\end{figure}

In Fig. \ref{fig:pdf-mu} we show the 
 PDF of the magnification. Note that 
it is clearly related to 
the convergence PDF since the 
two quantities,
related by
eq.~\eqref{magn}, agree up to a shift by one and a factor two within linear perturbation theory.

\begin{figure}
\centering
\includegraphics[width=\columnwidth]{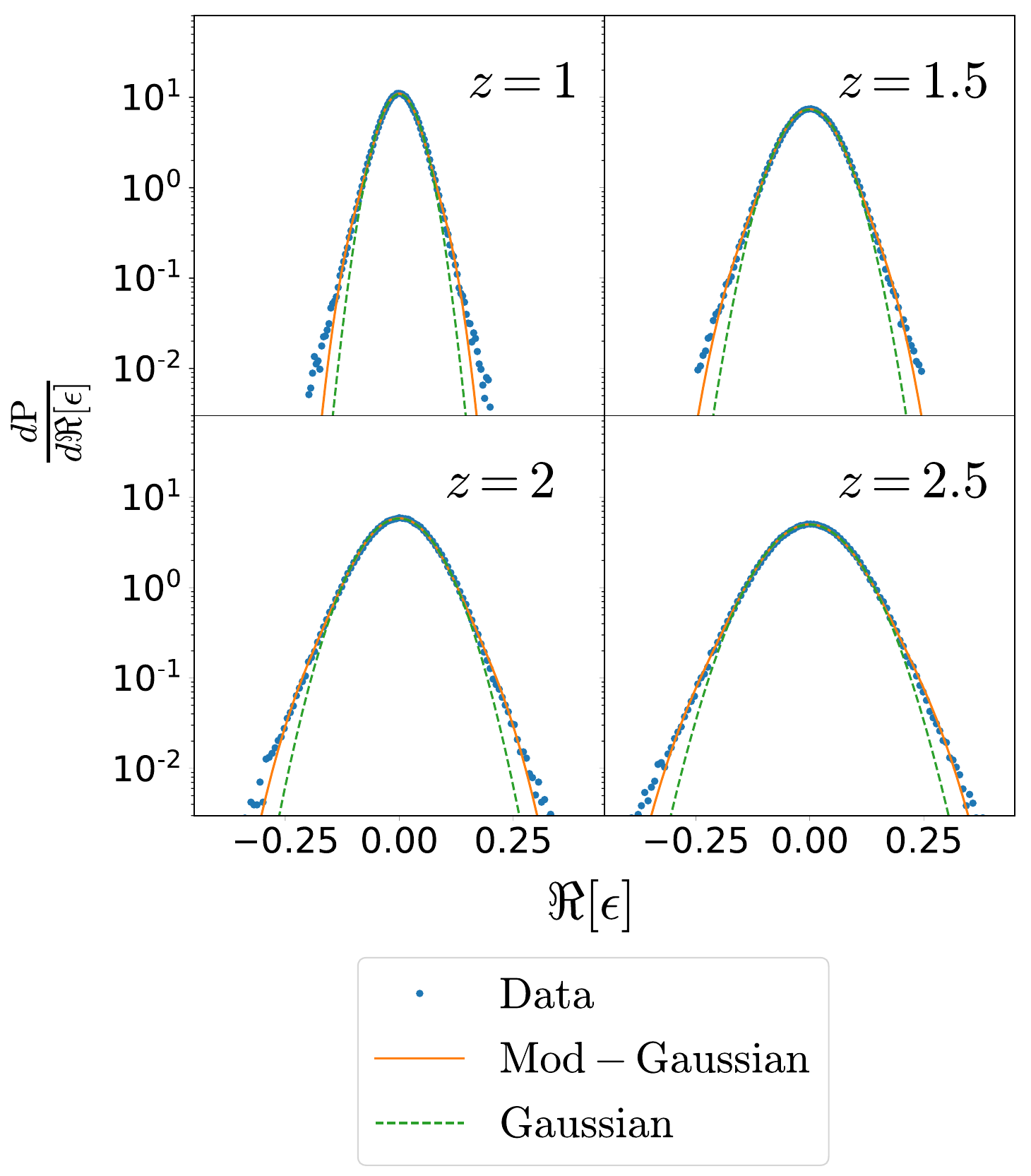}
\vspace{-10pt}
\caption{PDF for the real part of the ellipticity, compared to a Gaussian fit (green dashed lines) and to the modified Gaussian model given by eq.~(\ref{fit-mod-gauss}).}
\label{fig:pdf-Re-eps}
\end{figure}

In Fig. \ref{fig:pdf-Re-eps} we show the PDF
of the real part of the ellipticity,
$\Re[\ep] \equiv (\ep + \ep^\ast) / 2$.
In linear perturbation theory the convergence and the real part of the ellipticity have the same distribution.
However, non-linearities affect the two distributions in quite different ways: while $\ka$
is highly skewed (see Fig.\ \ref{fig:pdf-kappa-fit})
both the real and imaginary part of $\ep$
have the same symmetric, albeit non-Gaussian, distribution. Even though the power spectra of $\ka$ and $\ep/4$ are indistinguishable within our error bars, the PDF's of these quantities are clearly different. 
In Fig. \ref{fig:pdf-Re-eps} we compare the distribution for $\Re[\ep]$ to a Gaussian fit (green dashed lines). Close to the peak, at $\Re[\ep] = 0$, the PDF is well approximated by 
a normal distribution. However, the distribution has clearly visible, symmetric  non-Gaussian tails. 
The model for the probability distribution
$dP/d\Re[\ep]$ can be improved by
introducing a polynomial correction to the 
normal distribution of the form
\begin{equation}
\frac{dP}{d \Re[\ep]} = 
N_{\Re[\ep]}(1+ \alpha \Re[\ep]^6) \exp{\left(-\frac{\Re[\ep]^2}{2
\sigma^2_{\Re[\ep]}}\right)},
\label{fit-mod-gauss}
\end{equation}
where $N_{\Re[\ep]}$ is the 
normalisation, $\sigma_{\Re[\ep]}$
is the standard deviation of the 
Gaussian, and $\alpha$ is a parameter that models the non-Gaussian tail of the distribution.
The fit for the model in eq.~\eqref{fit-mod-gauss} is show in Fig. 
\ref{fig:pdf-Re-eps} (orange lines). 
The modified-Gaussian model
is a better fit than a simple normal distribution for large values of the 
ellipticity.
The values of the best-fit 
parameters are reported in Table \ref{table:eps}.
The parameter $\al$ decreases significantly with increasing redshift which is expected as more of the Gaussian signal from higher redshift is accumulated. At the same time, the average lensing signal and therefore $\si_{\Re[\ep]}$ increases with redshift.

\begin{table}
\begin{center}
\begin{tabular}{|c c c c| } 
 \hline
 $z$ & $\sigma_{|\ep|}$ & $\sigma_{\Re[\ep]}$
 & $\alpha$ \\ 
 \hline
 1   & 0.036  & 0.035 & $1.8 \times 10^6$ \\ 
 1.5   & 0.054 & 0.052 & $1.1 \times 10^5$\\ 
 2   & 0.068 &   0.066 & $2.1 \times 10^4$ \\ 
 2.5   & 0.079 & 0.078 & $7.0 \times 10^3$\\ 
 3   & 0.088 & 0.086 & $3.3 \times 10^3$\\ 
 \hline
\end{tabular}
\caption{
Fitted parameters for 
the $\Re[\ep]$-PDF and the $|\ep|$-PDF.
$\sigma_{|\epsilon|}$ is 
the standard deviation estimated from the fit of $|\ep|$ to a Rayleigh distribution,
 while $\sigma_{\Re[\ep]}$
and $\alpha$ are the parameters of the modified Gaussian PDF given in eqs.~\eqref{fit-mod-gauss} and \eqref{pdf-eps-mod}.
\label{table:eps}}
\end{center}
\end{table}

\begin{figure}
\centering
\includegraphics[width=\columnwidth]{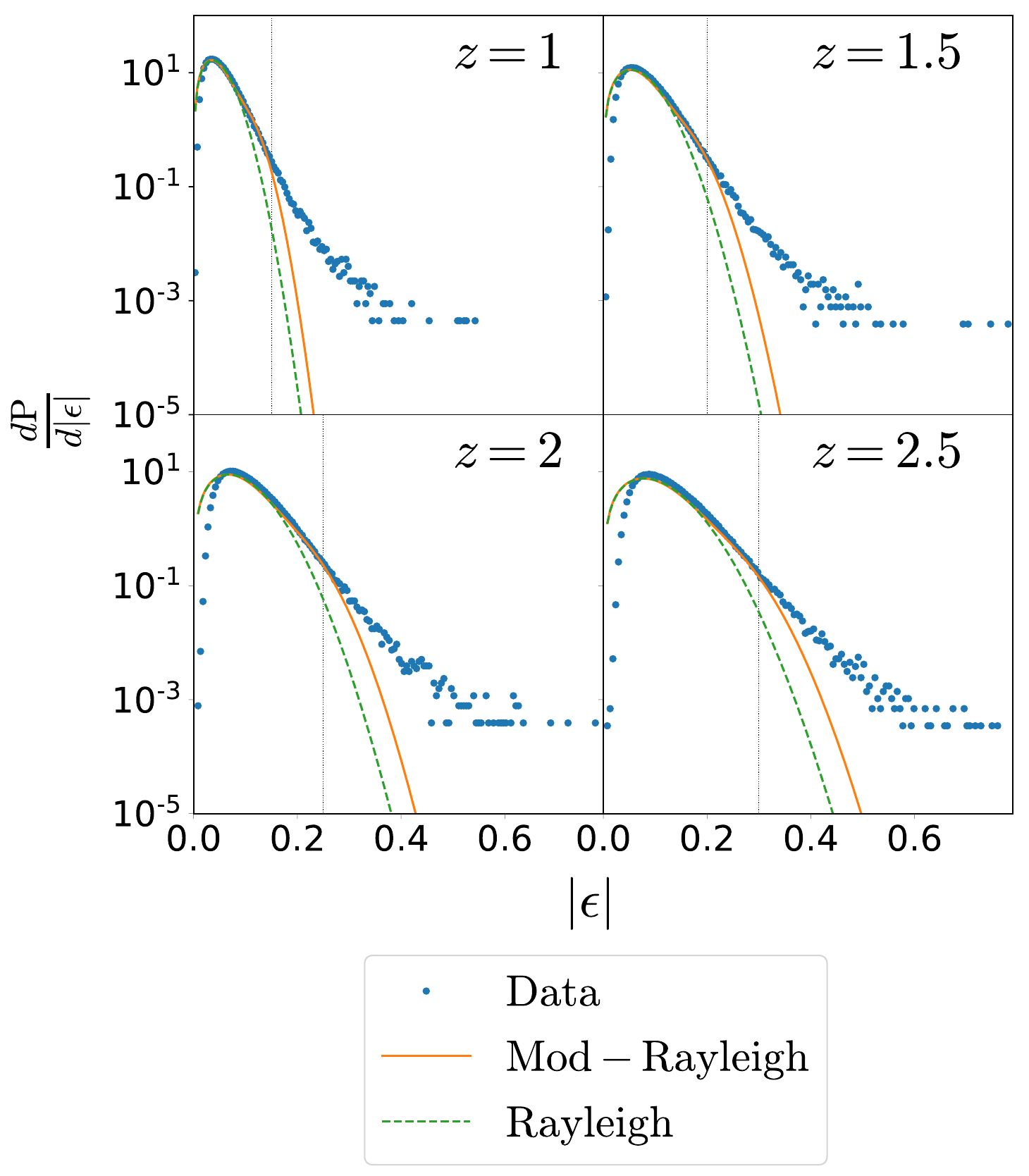}
\vspace{-10pt}
\caption{PDF for the estimated absolute value of the ellipticity for different redshifts. The orange solid lines show a fit for a modified Rayleigh distribution, whereas the green dashed lines show a fit for a standard Rayleigh distribution that is recovered if $\al = 0$ (see text for details). The vertical
dotted lines denote the maximum values of $|\ep|$ included in the fitted data.
}
\label{fig:pdf-eps-fit}
\end{figure}

Fig.\ \ref{fig:pdf-eps-fit} shows the probability distribution for the absolute value of the ellipticity, $|\ep|$. The PDF for $|\ep|$ is highly non-Gaussian already at high redshift.

Because of statistical isotropy, the joint distribution of $\Re[\ep]$ and $\Im[\ep]$ has to be a function of $|\ep| = \sqrt{\Re[\ep]^2+\Im[\ep]^2}$ only. We can therefore try to construct such an appropriate joint distribution given the knowledge of the fit for the marginalised distribution for $\Re[\ep]$. Integrating over the phase angle then directly determines the PDF for $|\ep|$. In the Gaussian case the latter is given by a Rayleigh distribution. These considerations motivate the following form of fitting function,
\begin{multline}
\frac{dP}{d|\epsilon|} = N_{|\epsilon|}
 \left[1-3 \alpha \left(\sigma_{\Re[\ep]}^6+\sigma_{\Re[\ep]}^4|\ep|^2 + \sigma_{\Re[\ep]}^2|\ep|^4\right) \right. \\
 + \left. \alpha |\ep|^6\right] 
  \frac{|\ep|}{\sigma^2_{\Re[\ep]}} \exp{\left(-\frac{|\epsilon|^2}{2 \sigma_{\Re[\ep]}^2}\right)}, \label{pdf-eps-mod}
\end{multline}
with  $\sigma_{\Re[\ep]}$ and $\alpha$ determined through a fit for $\Re[\ep]$, while $N_{|\ep|}$ is a normalization
parameter that we fit to the data.
Using the statistical isotropy argument it is easy to see that this ansatz is consistent with eq.\ \eqref{fit-mod-gauss}, while for $\al = 0$ the standard Rayleigh distribution is recovered.

In Fig.\ \ref{fig:pdf-eps-fit} we compare the PDF estimated
from our simulations to a fit to a Rayleigh distribution (green dashed lines) and to the 
modified Rayleigh distribution given in eq.~\eqref{pdf-eps-mod} (orange solid lines). 
The unmodified Rayleigh distribution 
provides a good fit only around
the peak of the distribution, while
there is a significant discrepancy
between data and model both 
for large value of $|\epsilon|$ 
and for $|\epsilon|$ approaching 
zero. 
The Rayleigh distribution underestimates the probability 
of having large values
of the ellipticity amplitude at all redshift. It compensates this somewhat by having a slightly larger width than the modified distribution, $\si_{|\ep|}>\si_{\Re[\ep]}$, but not very successfully. At high redshift the probability for $\Re[\ep]$ and $\Im[\ep]$
of being both close to zero
is significantly reduced compared to the Gaussian case. 
The modified Rayleigh distribution in eq.~\eqref{pdf-eps-mod}
is a much better fit to our data for large values of the ellipticity. However, for $|\ep| \rightarrow 0$, it reduces to
the Rayleigh distribution and therefore still overestimates the probability 
of having pixels with very small ellipticity.

\begin{figure}
\centering
\includegraphics[width=\columnwidth]{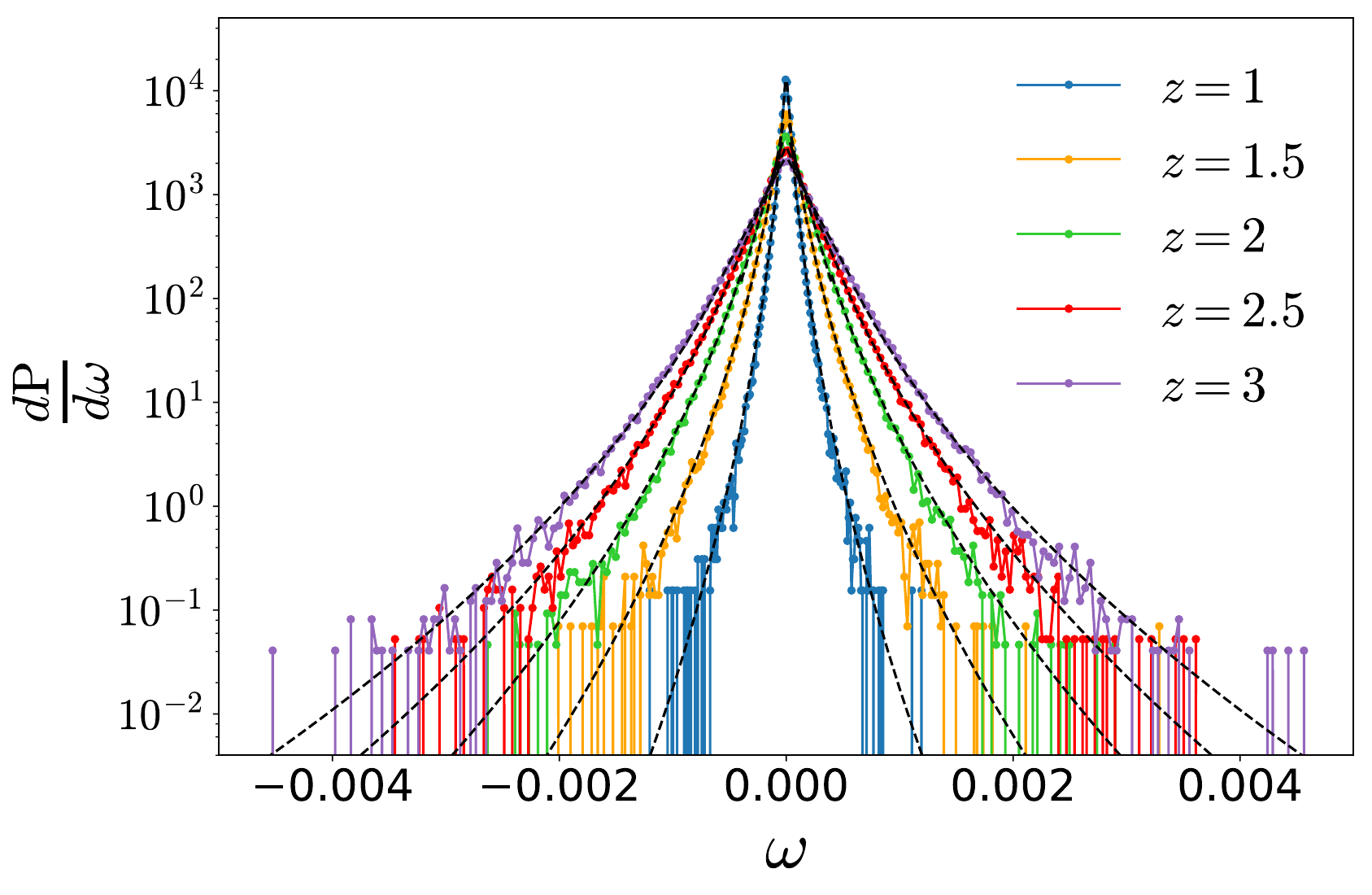}
\vspace{-10pt}
\caption{The PDF of the $\om$ (coloured) together with its fit (black dashed lines). The details of the fit are given in Appendix~\ref{omega-pdf}.}
\label{fig:pdf-omega-fit}
\end{figure}

In Fig. \ref{fig:pdf-omega-fit} we show the distribution
for the rotation angle $\omega$, which is not close to Gaussian at any redshift.
In Appendix~\ref{omega-pdf} we provide a derivation for  the expected PDF: for small values of $\omega$ it can be 
well modelled as an exponential, while the tails follow a product-Gumbel
distribution [see eq.~\eqref{e:extremevalue}] with the respective parameters $\lambda_m \sigma^2_\mathrm{lens}$ and $s$ given in Table~\ref{table:3}. 

\begin{table}
\begin{center}
\begin{tabular}{|c c c| } 
 \hline
 $z$ & $s$ & $\lambda_m \sigma^2_\mathrm{lens}$\\ 
 \hline
 1   & 0.00416554 &  0.00401273\\ 
 1.5   & 0.00562725 & 0.00600481\\ 
 2   & 0.00665294 &  0.00767290\\ 
 2.5   & 0.00763133& 0.00920118\\ 
 3   & 0.00855618 &  0.0102826\\ 
 \hline
\end{tabular}
\caption{Fitted parameters for the
$\omega$-PDF models discussed in Appendix 
\ref{omega-pdf}. The parameters $s$ represents the widths of the Gumbel distribution that characterises the tails, while $\lambda_m \sigma^2_\mathrm{lens}$ is the scale of the exponential near the maximum.\label{table:3}}
\end{center}
\end{table}

Note that here we show the probability distribution functions (PDFs) of the lensing quantities $\ka$, $\epsilon$ and $\omega$ while we can observe these quantities only at the position of a luminous source, where we, \eg can measure the ellipticity  of a galaxy. In our analysis we chose the source population to be drawn randomly from the $N$-body ensemble (thereby avoiding the additional complication of galaxy bias). What we obtain is therefore closer related to $\rho(z)\epsilon$ \textit{etc}., which means that the observables are mass-weighted in each pixel. Nevertheless, since the lensing quantities are integrated effects and get  contributions from density fluctuations along the entire light path, it is most probably a very good approximation to neglect their correlation with density fluctuations at the source redshift $z$ and our PDFs should be representative for the ones which are truly measured.

\subsection{Convergence from lensing potential}
\label{born}
As a consistency test for our results, we finally compare the angular power
spectrum and the map for the convergence
estimated using the full ray tracing method outlined in the previous sections (from $D_A$) and in the Born approximation, from the lensing potential map  (from $\Psi$). We described the method to obtain the convergence in the Born approximation in Sec.\ \ref{lens-pot-maps}.

\begin{figure}
\centering
\includegraphics[width=\columnwidth]{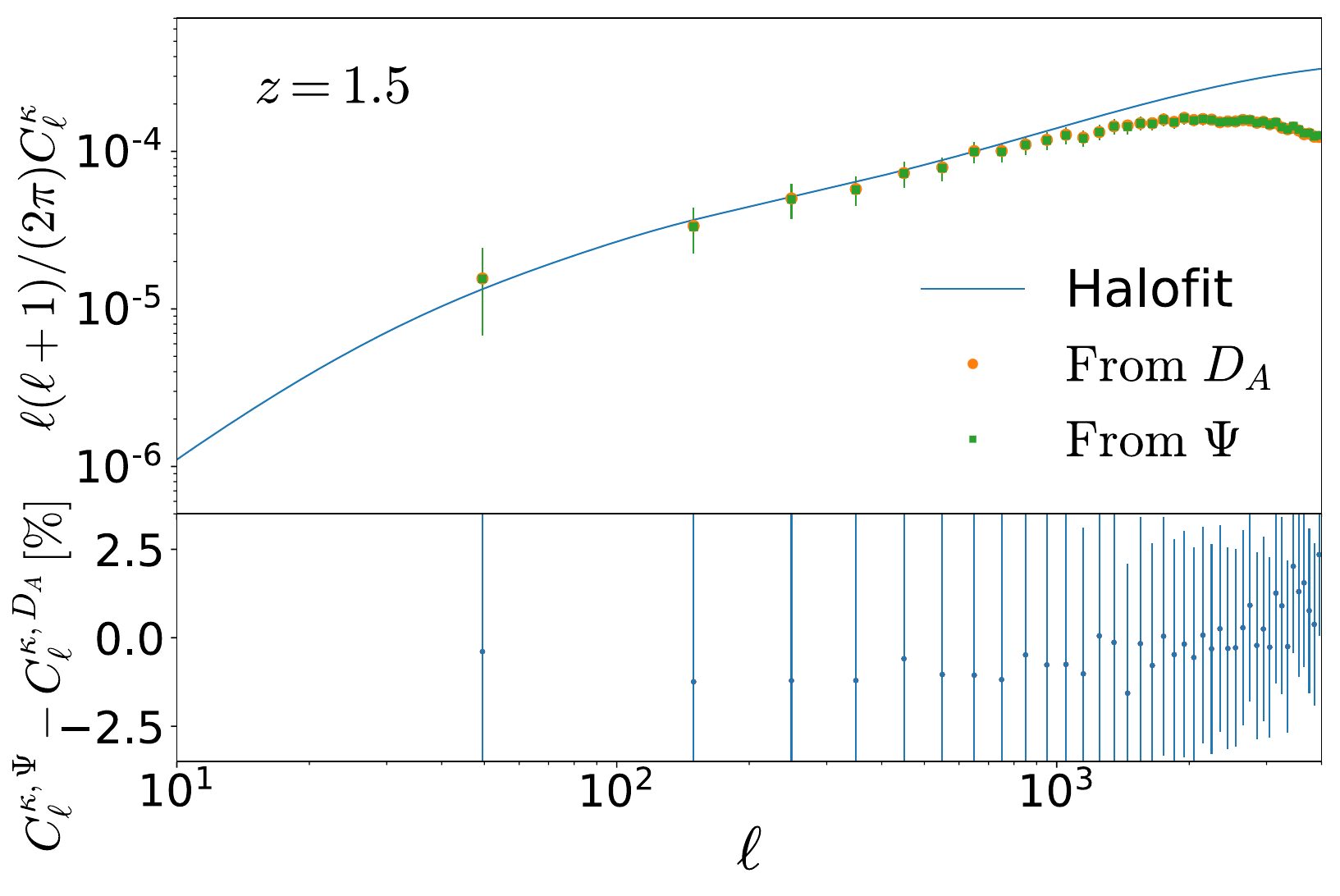}
\vspace{-10pt}
\caption{The angular power spectrum of the convergence obtained with the full ray tracing (\ie based on the non-perturbative area distance $D_A$) and
within the Born approximation (from the lensing potential $\Psi$). In the bottom panel we show the 
relative difference between the two spectra in \%, including error bars. }
\label{fig:cl-lens-pot}
\end{figure}

In Fig.~\ref{fig:cl-lens-pot} we show the 
angular power spectra of the convergence computed with the two methods. The methods agree within $2\%$. On scales $\ell \lesssim 1000$, where our 
computation is reliable, the Born approximation
systematically underestimates the power by about 1.5\%. 
However, it is not entirely clear whether the difference is physical or a computational artifact. We note in particular that the redshift selection is based on observed redshift for the ray tracer, while the method of the lensing potential assumes a distance--redshift relation.

\begin{figure}
    \centering
    \includegraphics[angle=90,origin=c,width=\columnwidth]{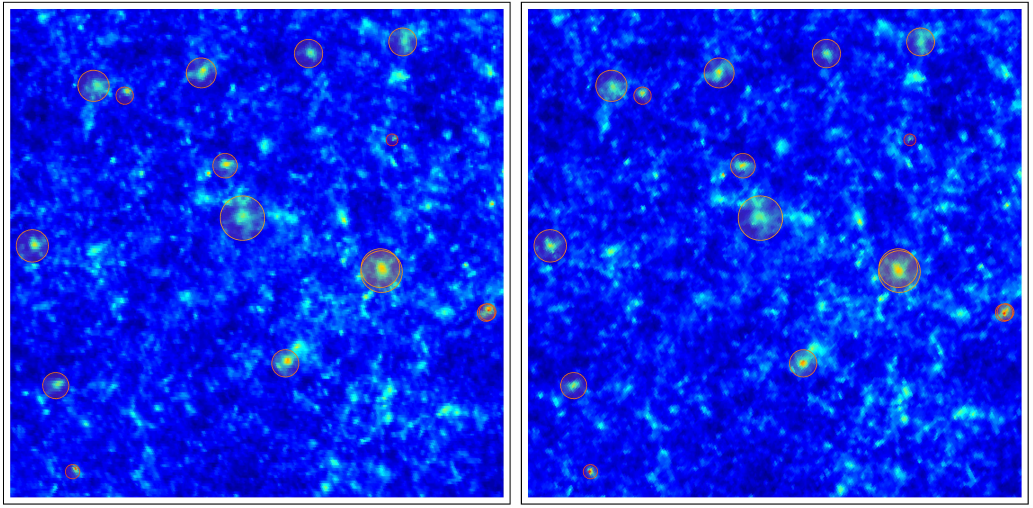}
    \caption{The upper panel shows a small patch of the $\ka$-map constructed from the observable properties of sources in the redshift bin at $z=1.5$ as computed with the ray tracer. The superimposed circles show the observed position and apparent size of dark matter halos with masses above $3\times 10^{14} M_\odot/h$ in the same field of view. The angular diameter subtended by the largest halo (close to the centre) is $\sim 18'$, while the smallest halo (towards top left) is only $\sim 5'$ across. For comparison, the lower panel shows the map of $\ka_\mathrm{lin}$ constructed from the lensing potential.}
    \label{fig:kappa_zoom}
\end{figure}

Fig.~\ref{fig:kappa_zoom} shows a zoom-in of the $\ka$-maps obtained with the ray tracer (top panel) and from the lensing potential (bottom panel) for the redshift bin at $z=1.5$. As visual reference points we also show the observed positions and apparent sizes of dark matter halos with more than $3\times 10^{14} M_\odot/h$ and observed redshifts below $1.5$. Here, the apparent size is inferred from the proper virial radius and the observed area distance $D_A$. While the agreement between the two methods of computing $\ka$ is generally excellent, one may notice at close inspection that the peaks of the $\ka$-map from the lensing potential do not line up perfectly with the positions of the clusters (which is the truly observed one in both panels). This error is due to the Born approximation, and we can estimate it to be of the order of the arcminute for this redshift.

\subsection{Image rotation}

\begin{figure}
    \centering
    \includegraphics[angle=90,origin=c,width=\columnwidth]{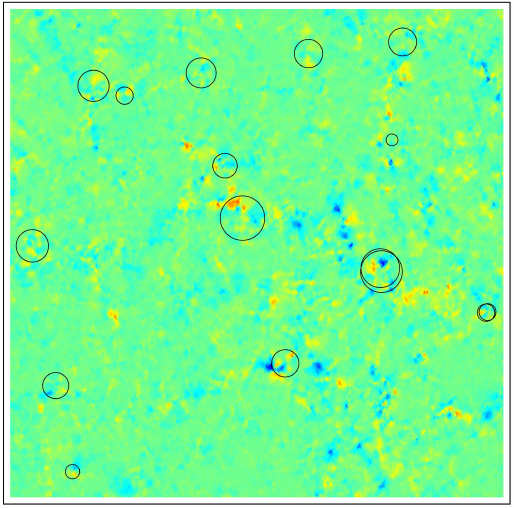}
    \caption{For the same patch as shown in Fig.~\ref{fig:kappa_zoom} we plot the map of the image rotation $\omega$ -- the colour scale ranges from $-0.002$ (blue) to $+0.002$ (red) radians.}
    \label{fig:omega_zoom}
\end{figure}

\begin{figure}
    \centering
    \includegraphics[width=\columnwidth]{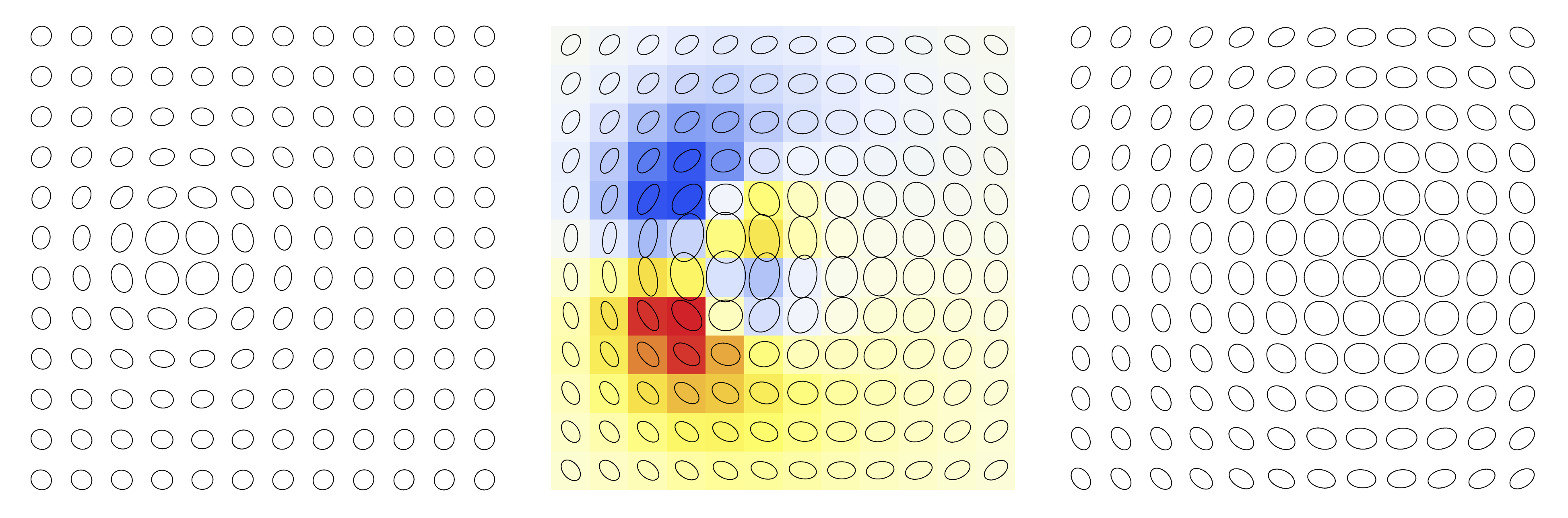}
    \caption{Illustration of the generation of image rotation through the composition of two linear lens maps. The left and right panels show the lens maps of two individual lenses, modelled here through rotationally symmetric Gaussian convergence profiles with misaligned centres of symmetry. The composition of the two maps is shown in the centre panel, and the colours indicate the image rotation (red and blue corresponding to opposite-sense rotation).}
    \label{fig:toyomega}
\end{figure}

Fig.~\ref{fig:omega_zoom} shows the map of the image rotation $\omega$ for the same redshift bin. The strongest rotation is not found at the centres of the projected dark matter halos, but typically in scattered places at the periphery of lensing peaks. The pattern can be understood quite well from considering a simplistic model where the rotation is generated by the composition of two linear lens maps, as illustrated in Fig.~\ref{fig:toyomega}. In the sketched example we place a background lens off-axis and close to the maximum of the shear distortion of a foreground lens. The composition of the two lens maps (done here in the Born approximation) has an anti-symmetric component, \ie some non-vanishing image rotation. In the shown configuration the rotation map has a parity-odd pattern with strong dipolar and noticeable quadrupolar components. Similar features (though less symmetric) can be seen in Fig.~\ref{fig:omega_zoom}.

\section{Conclusions}

In this paper we have presented the first study of weak lensing by cosmological structure using a high-resolution relativistic $N$-body simulation. To this end we have developed a methodology to solve the Sachs equations along the true photon trajectory. Our ray-tracing code is independent of the underlying cosmology. It only assumes that photons move along geodesics and works for arbitrary forms of the scale factor, the Bardeen potentials and the vector perturbations $B_i$, remaining completely agnostic about how these perturbations are generated. It is nonperturbative in the scalar part and perturbative to first order in the vector part. We have computed the observed positions and redshifts of a source population and the optical scalars $D_A$ (or rather $\ka=1-D_A/\bar D_A$), the ellipticity $\ep$ and the rotation $\om$ for the respective photon geodesics. We have also studied the angular one-point distributions of these variables and provide the first theoretically motivated fitting functions for the rotation.

Despite the fact that we nowhere invoke a Newtonian approximation, our results overall
agree well with previous ones obtained with Newtonian codes. Within the $\La$CDM cosmological standard model we therefore find no indication of large non-linear corrections coming from the gravity sector -- the non-linearity is completely dominated by the clustering of matter, while the gravitational fields remain weak in the Poisson gauge. The strong clustering of matter however means that the weak-lensing observables deviate significantly from linear perturbation theory in which $\om=0$ and both the $\ka$ and $\ep$ one-point distributions are Gaussian.
Our resolution has not been sufficient to extract the B-part of the ellipticity $\ep$ from the numerical noise.

While it is included in our results for the first time, we have not systematically studied the effect of the (small) vector perturbations which are induced at second order from the scalar perturbations and are completely absent in a Newtonian treatment. To see them
we would need to construct observables that specifically isolate them from other second-order contaminations.
We shall study this problem in future work.

\section*{Acknowledgements}
We thank Pierre Fleury and Enea Di Dio for insightful discussions and the authors of \citet{Hilbert:2019vca}, in particular Alexandre Barreira, for facilitating the code comparison shown in Appendix \ref{a:validation}.
This work was supported by a grant from the Swiss National Supercomputing Centre (CSCS)
under project ID s710. JA and CC acknowledge funding by STFC Consolidated Grant
ST/P000592/1. RD and FL thank the Swiss National Science Foundation for support.

This is a pre-copyedited, author-produced version of an article accepted for publication in Mon. Not. Roy. Astron. Soc. following peer review. The version of record, Monthly Notices of the Royal Astronomical Society Volume 497, Issue 2, September 2020, Pages 2078-2095, is available online at:
\url{https://doi.org/10.1093/mnras/staa2024}.

\paragraph*{Carbon footprint:} In this work we re-used existing simulation data. Additional post-processing consumed about $3000\, \mathrm{kWh}$ of electrical energy, which has an estimated\footnote{Our conversion factor of $0.2\, \mathrm{kg}\, \mathrm{CO}_2\, \mathrm{kWh}^{-1}$ is taken from \citet{VUARNOZ2018573}, table 2, assuming Swiss mix.} impact of $600\, \mathrm{kg}\, \mathrm{CO}_2$. This project also accounts for one return flight Geneva-London in economy class, causing emissions of approximately\footnote{ICAO Carbon Emissions Calculator, \url{https://www.icao.int/environmental-protection/CarbonOffset/Pages/default.aspx} --- retrieved 30.\ October 2019.} $180\, \mathrm{kg}\, \mathrm{CO}_2$.



\appendix
\section{General relation between the Jacobi map and the amplification matrix.}\label{a:A}

In this appendix we derive the relation between the Jacobi map and the amplification matrix, which is the Jacobian of the lens map, in full generality, \ie also for strong deflections.
We first introduce the deflection angle $\bal$ defined via the lens map, see Section~\ref{lens-pot-maps}, denoting the image position in our chosen coordinate system by $\bth_o=\bth(s_0)$ and the source position by $\bth_s=\bth(s)$, we set
\be \label{eA;defl}
 \theta^i(s) = \theta^i(s_0) +\al^i(s) \equiv (\bth(s_0) + \bal(s))^i \,.
\ee
Here $s$ is the affine parameter along the photon trajectory, $s_0$ is its value at the observer position and $\bth$ is the direction vector tracking the spatial photon position given in some coordinate basis $(\dd_0,\dd_i)$, and $\bal$ is the deflection angle. 
Note that the entire definition here is coordinate dependent and cannot be made intrinsic since
the quantities $\al^i(s)$, $\theta^i(s_0)$ and $\theta^{\prime i}(s)$ are best understood as coordinate vectors with respect to the coordinates $\dd_i$ and not as geometrical elements in a tangent space. In the given coordinates, 
the amplification matrix is now simply the Jacobian of this map, rescaled with the angular diameter distance of the background spacetime,
\be
{A^i}_{j}(s) = \bar D_A(s)\left(\de^i_{j} +{\al^i}_{,j}(s)\right) \,. 
\ee
Clearly, this construction makes sense only for lensing due to perturbations on some background, where the photon geodesic at $s$ and the observer position can be covered by one single coordinate patch.

To obtain the Jacobi map from ${A^i}_{j}$ we have to consider a neighbouring geodesic, say which has position $\bth(s_0) +\bep(s_0)$ at the observer and $\bth_s+\bep_s=\bth(s)+\bep(s)$ at the source in our chosen (arbitrary)  coordinate system. The Jacobi map expresses $\bep_s$ to first order, 
\be
(\bep(s))^i =\left(\ep^i(s_0) +\al^i_{,j}(s)\ep^j(s_0)\right) \,.
\ee
A Sachs basis at the observer $\bde_a(s_0)\,, a\in\{1,2\}$ is a 2d orthonormal basis of the plane (the ``screen'') normal to the photon direction $\bth(s_0)$ and the observer 4-velocity $u(s_0)$, and it is defined along the entire photon geodesic by parallel transport.
Therefore  $\bde_a(s)$ is unique up to a global rotation.
Let us denote by ${e^a}_i(s)$ the transformation from our coordinate basis $\dd_i$ to a fixed Sachs basis $\bde_a(s)$ and by  ${e^i}_a(s)$ its inverse, so that
the basis transformations are given by $\bde_a(s)={e^i}_a(s)\dd_i$ and $\dd_i={e^a}_i(s)\bde_a(s)$.

Our deviation vectors $\bep$ in the Sachs basis are
\begin{align}
\ep^{a}(s) &= {e^a}_i(s)\ep^{i}(s)\,,\\
\ep^{ a}(s_0) &= {e^a}_i(s_0)\ep^{i}(s_0)\,,\quad \ep^{ i}(s_0) = {e^i}_b(s_0)\ep^{b}(s_0) \qquad
\end{align}
so that
\be
(\bep)^a(s) =\left({e^a}_i(s){e^i}_b(s_0) + e^a_i(s)\al^i_{,j}(s)e^j_b(s_0)\right)\ep^b(s_0)\ \,.
\ee
which implies
\be\label{e:Jacob}
{D^a}_b(s) = \bar D_A\left({e^a}_i(s){e^i}_b(s_0) + {e^a}_i(s)\al^i_{,j}(s){e^j}_b(s_0)\right)\,.
\ee
 If two (say $\dd_1$ and $\dd_2$) of our arbitrary coordinates $\dd_i$ happen to be aligned with the Sachs basis and the deflection is sufficiently weak that we may consider $\bth \simeq \bn \simeq \bde_3$ constant (Born approximation), and the peculiar velocity is small, \ie approximately $u \propto \dd_0$, we can neglect the evolution of the Sachs basis and\footnote{More precisely, we must be able  to choose $\bde_i= f\dd_i$ where $f$ must be a normalisation which is constant along the geodesic.}
 $ {e^a}_i(s) \simeq \de^a_i$. In this case ${A^i}_j$ in the 2d space normal to $\bn$ agrees with ${D^a}_b$.  Interestingly, in Poisson gauge this is the case at first order in perturbation theory if there are only scalar perturbations. The main reason for this is that the  spatial part of the metric in Poisson gauge is conformally flat and parallel transport does not rotate the Sachs basis along the photon geodesic [see~\citet{DiDio:2019rfy} for details].
The opposite extreme case is geodesic light-cone gauge~\citep{Gasperini:2011us}, where the angular coordinates along a photon geodesic are chosen to be its angular position at the observer so that, by definition $\bal\equiv 0$. In these coordinates the amplification matrix is proportional to the identity and the Jacobi map is given by the first term of \eqref{e:Jacob}. In this case the metric is far from conformally flat and parallel transport rotates the Sachs basis along the photpon geodesic so that ${e^a}_i(s)$ has a non-trivial $s$-dependence. Of course for a fixed Sachs basis at the observer, $\bde_a(s_0)$ one finds  the identical Jacobi map in both coordinate systems.

In general, the Sachs basis varies along the line of sight and ${D^a}_b$ is not of the simple form of a gradient map like the Jacobian of eq.~\eqref{e:gradmap}. This then implies that the power spectrum of the rotation and the B-mode of the Jacobi map do in general not agree. \citet{Hirata:2003ka} showed that for a gradient map, ${A^i}_j(\bth) = {\rm const} + {\al^i}_{,j}(\bth)$ in the flat sky approximation, we always have $C_\ell^\ka=C_\ell^{\gga_E}$ and $C_\ell^\om=C_\ell^{\gga_B} =0$.
Since the Jacobi map agrees with the coordinate-dependent amplification matrix only at first order in perturbation theory (Born approximation) in Poisson gauge, these relations are not expected to hold beyond first order perturbation theory. 

\section{Shear and image rotation at leading order}
\label{omega-pdf}

In this appendix we derive some expressions for the ellipticity $\ep$ and the rotation $\om$ at leading order. For this purpose it is sufficient to work in the Born approximation $(n^i \simeq \mathrm{constant})$. We can also neglect Shapiro delays and set $d\tau = -dr$, with $r$ denoting the comoving distance from the observer.

Let us begin by inspecting eq.~(\ref{e:Sachssi}) which at lowest order can be approximated as
\be \label{e:app1}
\frac{d\tilde{\si}}{d\tau} \tilde{D}_A^{-2} \simeq -\left(\tilde{e}^i_1 \tilde{e}^j_1 - \tilde{e}^i_2 \tilde{e}^j_2 + \ii \tilde{e}^i_1 \tilde{e}^j_2 + \ii \tilde{e}^i_2 \tilde{e}^j_1\right) \partial_i \partial_j \frac{\phi+\psi}{2} \equiv \tilde{\psi}_0\,,
\ee
where we drop contributions from $B_i$ which is only generated at second order in $\Lambda$CDM cosmology. We introduce $\tilde{\psi}_0$ as a shorthand for the complex source term that is generated by Weyl curvature.

Since $\tilde{\si}$ and $\tilde{\psi}_0$ are first-order quantities, we only need the background solution of $\tilde{D}_A$ which is given by $\tilde{D}_A \simeq r$ or, equivalently, $D_A \simeq r / (1+z)$. Therefore
\be
\tilde{\si} \simeq -\int\limits_0^r (r')^2 \tilde{\psi}_0(r') dr'\,.
\ee

We now continue with the same reasoning by approximating eqs.~(\ref{e:ellipom}) at leading order as
\be
\frac{d\ep}{d\tau} \simeq 4 \frac{\tilde{\si}}{r^2}\,, \qquad \frac{d\om}{d\tau} \simeq \frac{\ii \left(\ep^\ast \tilde{\si} - \ep \tilde{\si}^\ast\right)}{8 r^2} \,,
\ee
so that by inserting the previous result we get an estimate of the complex ellipticity of
\be
\ep \simeq \int\limits_0^r \frac{4 dr_1}{r_1^2} \int\limits_0^{r_1} r_2^2 \tilde{\psi}_0(r_2) dr_2 = 4 \!\int\limits_0^r \left(1-\frac{r'}{r}\right) r' \tilde{\psi}_0(r') dr'\,.
\ee
For the rotation $\om$ we then get
\begin{multline}
 \om \simeq \int\limits_0^r \frac{\ii\, dr'}{2 (r')^2} \int\limits_0^{r'} dr_1 \int\limits_0^{r'} dr_2 r_1^2 \left(1-\frac{r_2}{r'}\right) r_2\\ \times\left[\tilde{\psi}_0^\ast(r_1)\tilde{\psi}_0(r_2) - \tilde{\psi}_0(r_1)\tilde{\psi}_0^\ast(r_2)\right]\,,
\end{multline}
which, after some manipulation, can be rearranged to
\begin{multline}\label{e:ompert}
\om \simeq \int\limits_0^r \int\limits_0^r dr_1 dr_2 \Re[\tilde{\psi}_0(r_1)] \Im[\tilde{\psi}_0(r_2)] \\ \times r_1 r_2 (r_1 - r_2) \left(\frac{1}{r} - \frac{1}{\mathrm{max}(r_1,r_2)}\right)\,.
\end{multline}
We observe that $\om$ is given by a double integral of $\tilde{\psi}_0$ with a kernel that vanishes in the coincident limit and generates maximum signal for a pair of lenses that are placed at one third and two thirds of the comoving distance between source and observer. More precisely, the kernel takes its maximum value, $r^2/27$, at $r_1 = r/3$, $r_2 = 2r/3$, and is parity-odd under $r_1 \leftrightarrow r_2$.

The last expression can be used, for instance, to estimate the two-point function of $\om$, which then depends on the first four $n$-point functions of the gravitational potentials $\phi$ and $\psi$. We can also get some insights into the one-point distribution of $\om$ that we discuss next.

Due to the fact that the kernel vanishes for $r_1 = r_2$, large values of $\om$ can only be generated by pairs of gravitational lenses that are well separated along the line of sight. We can therefore assume that these lenses are uncorrelated to a good approximation. Furthermore, if we neglect cosmological evolution, statistical isotropy implies that $\Re[\tilde{\psi}_0]$ and $\Im[\tilde{\psi}_0]$ are drawn from the same distribution.

Let us now first consider the simplest case of two lens planes, obtained by splitting the line of sight in half and treating the source terms on both legs as uncorrelated random variables. A rough estimate of $\om$ is then given by
\be
\om \simeq \frac{r^4}{192} \left(X_1 Y_2 - X_2 Y_1\right)\,,
\ee
where $X_j$, $Y_j$ denote the random variables on lens plane $j$, respectively given by the real and imaginary parts of its ``effective'' source term $\tilde{\psi}_0$. The factor $1/192$ comes from integrating the kernel, approximating $\tilde{\psi}_0$ by these piecewise constant ``effective'' values.

In the case where the $X_j$, $Y_j$ are Gaussian with r.m.s.\ amplitude $\si_\mathrm{lens}$, we can compute the probability density function of $\om$ explicitly,
\be
    p(\om) = \frac{96}{r^4 \si_\mathrm{lens}^2} \exp\!\left(-\frac{192}{r^4 \si_\mathrm{lens}^2}\vert\om\vert\right)\,.
\ee
Note that $\si_\mathrm{lens}$ is the typical amplitude of the effective $\tilde{\psi}_0$ and therefore has dimensions of inverse area. The above expression would suggest that the r.m.s.\ value of $\om$ scales as $\sim r^4$, but this is too naive because $\si_\mathrm{lens}$ would depend on the effective baseline that contributes to each lens plane.

We can try to resolve this issue by considering multiple lens planes spaced regularly along the line of sight. Their separation can then be chosen to be commensurate with the correlation length of $\tilde{\psi}_0$, so that the planes can still be considered approximately uncorrelated. In this case $\si_\mathrm{lens}$ should be approximately constant and can be estimated as the typical scale of the gravitational potential ($\sim 10^{-5}$ in cosmological settings) divided by the square of a characteristic length scale, expected to be similar to the correlation length of $\tilde{\psi}_0$. In this approximation of multiple lens planes with random $X_j$, $Y_j$ we get
\be\label{e:multilensplane}
\om \simeq \sum_{j,k}^n A_{jk} X_j Y_k\,,
\ee
where $n$ is the number of lens planes and the coefficient matrix $A_{jk}$ is computed from eq.~(\ref{e:ompert}) assuming that $\tilde{\psi}_0$ is approximately piecewise constant,
\be\label{e:coeffmatrix}
 A_{jk} \equiv \!\int\limits_{r_j-\frac{\Delta r}{2}}^{r_j+\frac{\Delta r}{2}} \!dr_1\! \int\limits_{r_k-\frac{\Delta r}{2}}^{r_k+\frac{\Delta r}{2}} \!dr_2 r_1 r_2 (r_1 - r_2) \left(\frac{1}{r} - \frac{1}{\mathrm{max}(r_1,r_2)}\right)\,.
\ee
In this expression, $r_j$ denotes the comoving distance to lens plane $j$, and $\Delta r = r/n$ is the length of the distance interval assigned to each lens plane.

In the Gaussian case the probability density function of $\om$ can be calculated explicitly, assuming for simplicity that the $X_j$, $Y_k$ are all independent and identically distributed\footnote{It would be possible to take into account cosmological evolution by absorbing the differential amplitudes of different lens planes into the coefficient matrix $A_{jk}$.} with r.m.s.\ amplitude $\si_\mathrm{lens}$,
\begin{multline}
p(\om) = \int \frac{d^nX d^nY}{\left(2\pi \si_\mathrm{lens}^2\right)^n} \exp\left(-\frac{\sum_j^n X_j^2 + \sum_k^n Y_k^2}{2 \si_\mathrm{lens}^2}\right)\\ \times \delta\left(\om - \sum_{j,k}^n A_{jk} X_j Y_k\right)\,.
\end{multline}

The Gaussian integrals can be solved by defining a $2n$-dimensional random vector $Z \equiv (X_1, \ldots, X_n, Y_1, \ldots, Y_n)$ and a $2n\times2n$-matrix $M$ as
\be
    M \equiv \frac{1}{2} \begin{pmatrix}
0 & A \\
A^T & 0
\end{pmatrix}\,,
\ee
such that
\begin{multline}
p(\om) = \int \frac{d^{2n}Z}{\left(2\pi \si_\mathrm{lens}^2\right)^n} \exp\left(-\frac{\sum_j^{2n} Z_j^2}{2 \si_\mathrm{lens}^2}\right)\\ \times\delta\left(\om - \sum_{j,k}^{2n} M_{jk} Z_j Z_k\right)\,.
\end{multline}
Next we make a change of variables to rotate into the orthonormal eigenbasis of $M$. Due to the symmetries of $M$, all its eigenvalues have multiplicity $2$, and for each eigenvalue $\lambda$, $-\lambda$ is also an eigenvalue due to the antisymmetry of $A$. The Gaussian integrals can then be solved recursively by working through the list of distinct positive eigenvalues of $M$. One obtains
\be
 p(\om) = \!\sum_{\lambda_m > 0} \frac{1}{4 \lambda_m \si_\mathrm{lens}^2} \exp\left(\!-\frac{\vert\om\vert}{2 \lambda_m \si_\mathrm{lens}^2}\right) \prod_{\substack{\lambda_{m'} > 0\\ m' \neq m}} \frac{\lambda_m^2}{\lambda_m^2 - \lambda_{m'}^2}\,.
\ee
Therefore, the variance of $\om$ is given by
\be
 \langle\om^2\rangle = 8 \si_\mathrm{lens}^4 \sum_{\lambda_m > 0} \lambda_m^2 \prod_{\substack{\lambda_{m'} > 0\\ m' \neq m}} \frac{\lambda_m^2}{\lambda_m^2 - \lambda_{m'}^2}\,.
\ee
The tail of the distribution is dominated by the largest eigenvalue $\lambda_\mathrm{max}$, and the asymptotic behaviour is $p(\om) \sim \exp(-\vert\om\vert / 2 \lambda_\mathrm{max} \si_\mathrm{lens}^2)$. With $A_{jk}$ given by eq.~(\ref{e:coeffmatrix}) we find that $\lambda_\mathrm{max} \sim r^4 / n$, which means that for $\Delta r = r/n$ fixed, the r.m.s.\ amplitude of $\om$ should scale as $\sim r^3$.

In the real Universe matters are more complicated because $\tilde{\psi}_0$ is poorly described by a Gaussian field. In particular, the distribution of $\tilde{\psi}_0$ has a significant non-Gaussian tail of strong lenses. In this situation it may be justified to assume that extreme values of $\om$ are dominated by those lines of sight which encounter two separate extreme values of $\tilde{\psi}_0$. We can then use extreme value theory to make statements about the tails of the probability distribution of $\om$.

Let us be very simplistic here and neglect the coefficient matrix in eq.~(\ref{e:multilensplane}) and pretend that instead $\om$ is given by a sum of products of independent and identically distributed pairs of random variables, not necessarily Gaussian. If, as we said, for large $\om$ this sum is dominated by a single term, generated from the product of the maximum value $X_\mathrm{max}$ of all the $X_j$ with the maximum value $Y_\mathrm{max}$ of all the $Y_k$, the probability density of $\om$ becomes a product distribution based on the distribution of those maximum values.

The Fisher--Tippett--Gnedenko extreme value theorem states that for a large number of draws $X_j$ from a realistic and possibly non-Gaussian distribution,
the distribution of $X_\mathrm{max}$ asymptotically approaches the Gumbel distribution (and likewise for $Y_\mathrm{max}$). Since this distribution is known, we can again estimate the distribution of $\om$ from
that extreme pair. We get
\begin{multline}\label{e:extremevalue}
p(\om) \sim \int_{-\infty}^\infty \frac{dX_\mathrm{max}}{s^2 \vert X_\mathrm{max}\vert}\exp\biggl(-\frac{X_\mathrm{max} - 2 \mu + \vert\om\vert/X_\mathrm{max}}{s}\biggr.\\ \biggl.- \er^{-(X_\mathrm{max}-\mu)/s} - \er^{-(\vert\om\vert/X_\mathrm{max} - \mu)/s}\biggr)\,,
\end{multline}
with two unknown parameters $s$ and $\mu$ that depend on the distribution of $X_j$. The asymptotic behaviour can be studied numerically and the parameters can be determined by fitting to observations.

In practice we expect that the above expression is a good description only in the tails of $p(\om)$. To find the best fit, we consider only the bins outside the $95\%$ region centered on $\om=0$. For such large values of $\vert\om\vert$ the asymptotic distribution (\ref{e:extremevalue}) is dominated by the exponential behaviour of the Gumbel distribution for large $X_\mathrm{max}$, and $\mu$ is therefore nearly degenerate with the normalisation. This allows us to set $\mu \simeq 0$ and fit only for $s$ and the normalisation. We verified that treating $\mu$ as a free parameter does not improve the goodness of fit noticeably.

\section{Pixelisation and resolution error}
\label{res-test}
The angular power spectra extracted from 
our simulations exhibit a suppression of power
on small scales, reaching $\sim 50\%$ at $\ell = 3000$. 
We identify two sources of error introduced by our method that cause a small-scale power suppression: the pixelisation of the fields 
and the finite resolution of the simulations.
In this section we quantify how much these
two effects impact our results. 

The pixelisation error is due to the fact 
that we estimate the value of a field (\ie $\kappa, \ep$ or $\omega$) in a pixel by averaging over the observed values at the particle positions for all particles whose observed position falls within that pixel. In each pixel we have a different number of particles and the mean of
their positions does not correspond to the centre
of our pixel. 
In order to quantify the error introduced by
the pixelisation, we generate a high resolution Gaussian map for the convergence ($N_\text{side} = 8192$) and we smooth this map into a lower resolution map in the following way:
the value of the convergence in each pixel of the smoothed map is computed by averaging over 
a subset of the high-resolution map values within that pixel. Each value of the convergence within the pixel has a $30\%$
probability of being accepted, which yields a sample rate similar to low-density regions in our source catalogue. We performed this analysis for 
three different smoothed maps, with 
$N_\mathrm{side} = 512, 1024, 2048$. 

\begin{figure}
\centering
\includegraphics[width=\columnwidth]{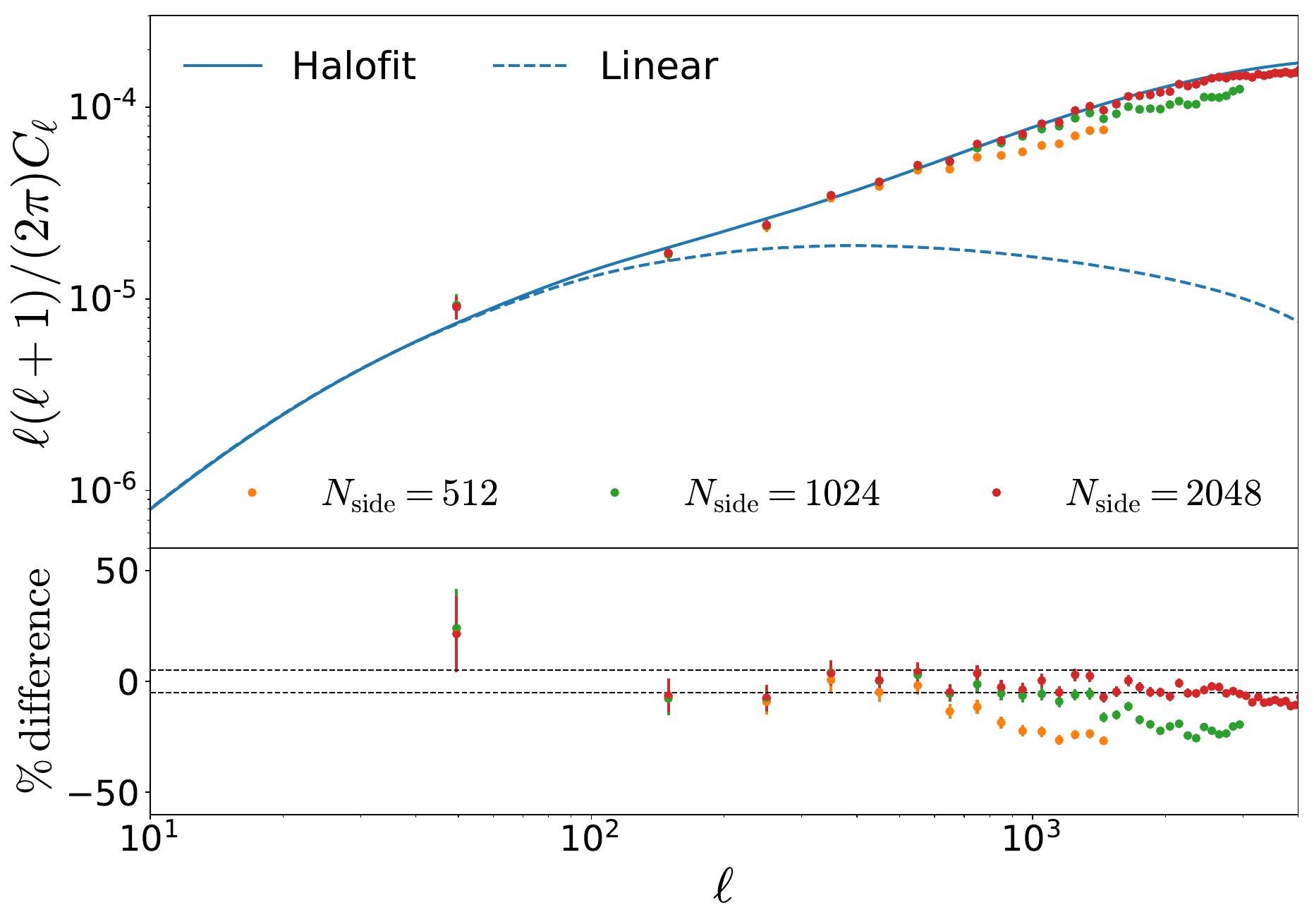}
\vspace{-10pt}
\caption{Impact of pixelisation on the convergence spectrum for $N_\mathrm{side} = 512, 1024, 2048$. 
The blue line denotes the prediction from
linear theory (dashed line) and \textsc{Halofit} (solid line).}
\label{fig:pixelisation}
\end{figure}

In Fig.~\ref{fig:pixelisation} we show 
the angular power spectra computed for the 
three simulated smoothed maps and we compare the results with the \textsc{Halofit} prediction for the convergence spectrum. The pixelisation suppresses power on small scales
as expected, and the effect depends on the number of pixels ($N_\mathrm{pixel} = 12 \, N^2_\text{side}$). For $N_\text{side} = 2048$, the power suppression is $< 5\%$ for $\ell\lesssim 3000$. Therefore, we 
adopt this angular resolution for the analysis presented in the main part of this paper.

Another cause of small-scale power suppression
is the finite grid resolution of our simulation.
In fact, \emph{gevolution} is a particle-mesh code and therefore modes smaller than 
the Nyquist frequency cannot be resolved. 
The resolution of the grid in our simulation
is $\Delta r = 312.5\, \mathrm{kpc}/h$,
which correspond to a Nyquist frequency 
$k_\mathrm{Nyq} \approx 10 \,h/\mathrm{Mpc}$.
This resolution allows us to resolve angular scales of around $\ell_\mathrm{max} \sim r(z)k_\mathrm{Nyq} \sim 20000$
at $z = 1$. 
However, the weak-lensing observables are
integrated fields. Therefore, the larger resolution
error from smaller redshifts propagates into
the estimated convergence at all redshifts.

In order to quantify this effect, we 
model the power suppression in the power spectrum 
as
\be
P_\text{sim}(k) \approx P_\text{true}(k)\left[1 -
A \left(\frac{k}{k_\text{Nyq}}\right)^2\right],
\label{Ps-res}
\ee
where $P_\text{sim}$ is the power spectrum 
in our particle-mesh simulation, while 
$P_\text{true}$ denotes the exact non-linear power spectrum in the continuum limit. $A$ is a smooth function of redshift that can be determined empirically. 

In order to quantify the impact of the suppression
in eq.~\eqref{Ps-res} on the convergence and
shear spectrum, we compute the 
convergence spectrum in the Limber approximation~\citep{Kaiser:1992ps,Scoccimarro:1999kp, Lemos:2017arq}
\begin{multline}
C^{\kappa}_\ell(z) = \frac{9}{4} \left(\frac{H_0}{c}\right)^4 \Omega_\text{m}^2
\int^{r(z)}_0 dr
\left[\frac{D_{+}(a)}{a} \frac{(r(z) - r)}{r(z)}\right]^2\times \\
P_\text{sim}\left(\frac{\ell + 1/2}{r}, z = 0 \right),
\label{Clres}
\end{multline}
where $r(z)$ is the comoving distance
at the source redshift (i.e. the mean redshift in the bin), $D_{+}(a)$ is the linear growth factor normalised to 1 at $z = 0$ and 
$P_\text{sim}$ is computed from eq.~\eqref{Ps-res}
assuming the \textsc{Halofit} power spectrum to be
the true power spectrum. In eq.\ \eqref{Clres} we neglect the scale-dependence of the growth for simplicity. However, this approximation does not significantly impact the result of our error analysis where our model for the power suppression in eq.\ \eqref{Ps-res} is mainly valid at large scales.

\begin{figure}
\centering
\includegraphics[width=\columnwidth]{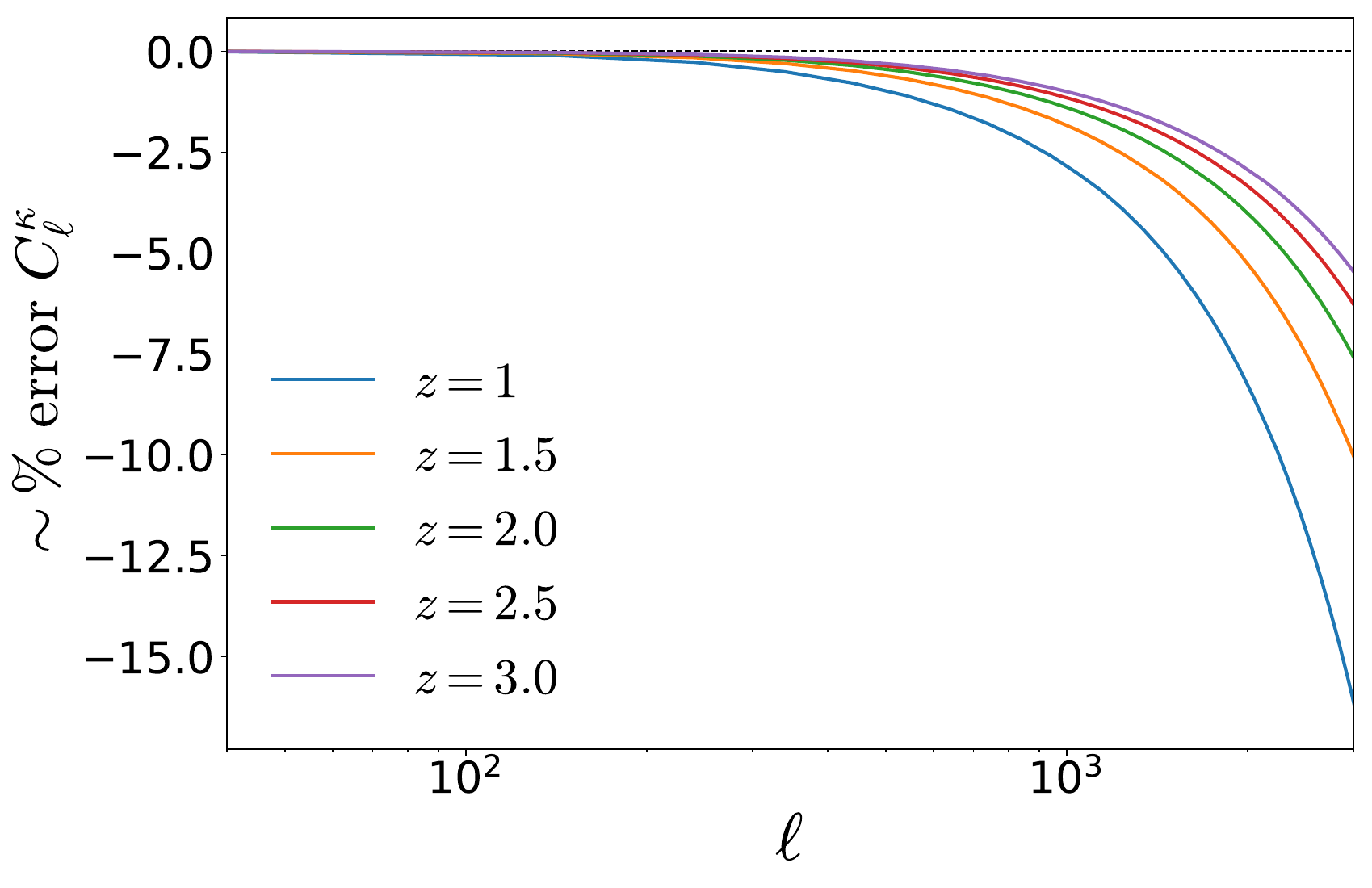}
\vspace{-10pt}
\caption{The relative difference between $C_\ell^\ka$ from the fully resolved \textsc{Halofit} power spectrum and from the power spectrum suppressed on wave numbers above $k_\text{Nyq}=10 \,h/\mathrm{Mpc}$.}
\label{fig:res0}
\end{figure}

In Fig. \ref{fig:res0} we show the relative difference between the suppressed power spectrum 
in eq.~(\ref{Clres}) and the angular power spectrum for an ideal simulation with $k_\text{Nyq} \rightarrow \infty$. The factor $A$ has been set to 1 in this plot. 

\begin{figure}
\centering
\includegraphics[width=\columnwidth]{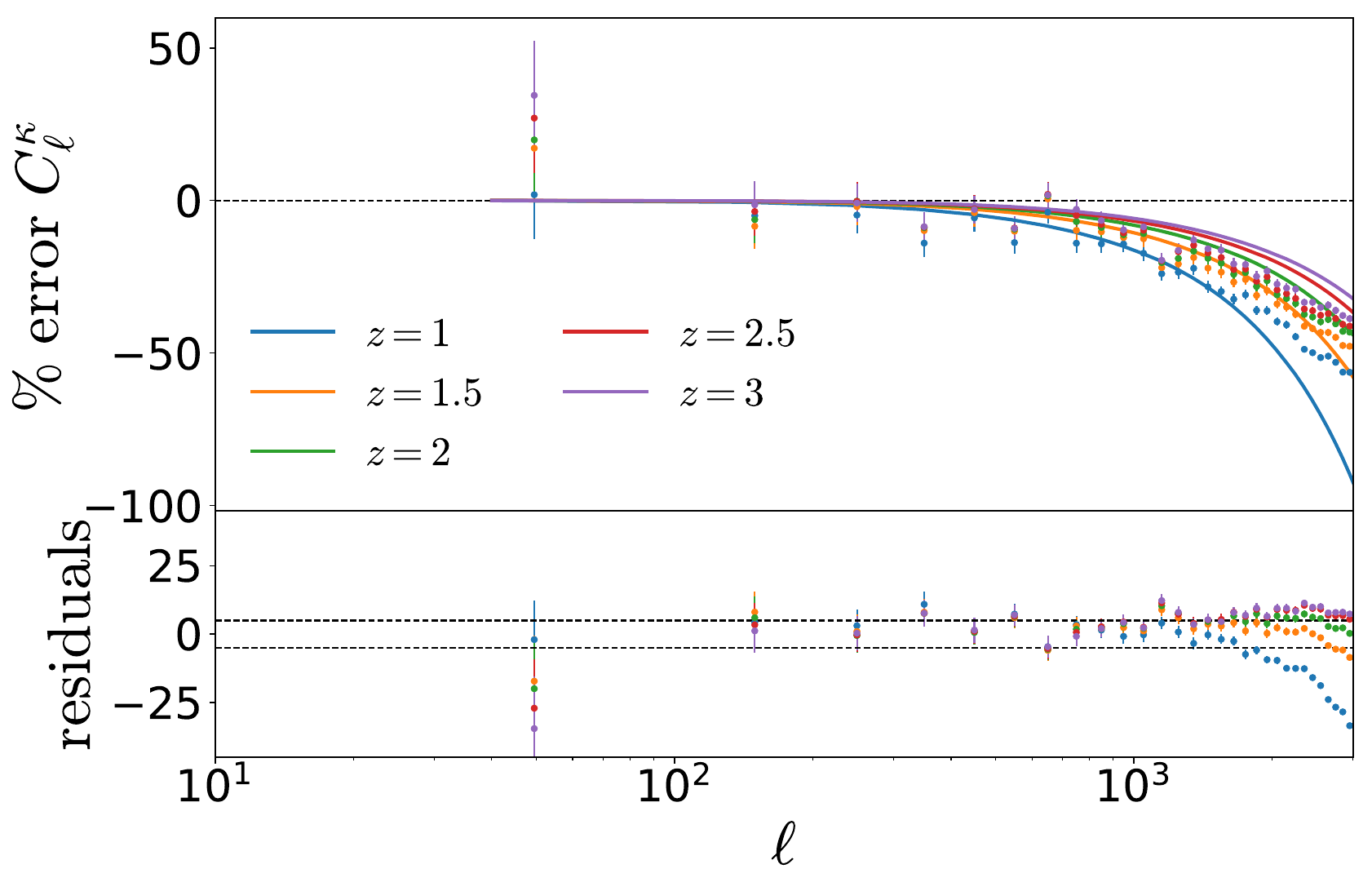}
\vspace{-10pt}
\caption{The convergence power spectrum from our simulation compared to the simple power suppression model given in eqs.~\eqref{Ps-res} and \eqref{Clres}.}
\label{fig:res1}
\end{figure}

The factor $A$ in eq.~\eqref{Ps-res} can be estimated directly from the simulation power spectra. 
The power spectra computed from
\emph{gevolution} show a drop of power 
on small scales due to the finite resolution
of the grid during the gravitational evolution, which effectively leads to a softening of the gravitational forces. 
Assuming that the 
``true'' power spectrum is given by the \textsc{Halofit} recipe we can get an empirical fit of $A$ due to modified gravitational evolution. 

In addition, since the gravitational potential in the simulation is computed from a smoothed density field, \ie the density field is estimated through cloud-in-cell (CIC) particle to mesh projection, there is another contribution to the coefficient $A$ due to this projection.
In the estimation of the matter power spectrum, the CIC kernel is deconvolved
and therefore already accounted for. This correction obscures the fact that the CIC projection still has an effect on the gravitational potential. In other words, the coefficient $A$ has two components, $A = A_\text{evo} + A_\text{CIC}$, where $A_\text{evo}$ describes the empirical suppression (due to force softening) of the matter power spectrum as explained above, and the
coefficient $A_\text{CIC}$ can be estimated analytically from the CIC smoothing kernel as $A_\text{CIC} = \pi^2/12$.

Taking into account both the finite-resolution loss of power and the cloud-in-cell smoothing, we estimate the coefficient $A$ to be in the range $A\in [4.5, 6]$ for redshifts in the range $z\in [1, 3]$. 

In Fig.~\ref{fig:res1} we compare the relative difference between the convergence spectrum estimated by our method 
and the \textsc{Halofit} power spectrum and the
model for the power-loss given by
eq.~\eqref{Ps-res}.
Up to $\ell \approx 2000$, the simple model in eqs.~\eqref{Ps-res} and \eqref{Clres} explains 
the small-scale power suppression in the weak-lensing observables obtained from our simulation.

\section{Validation of the lensing potential calculation}
\label{a:validation}

In order to validate our method to compute $\ka_\mathrm{lin}$ directly in pixel space from the lensing potential (see Sec.\ \ref{lens-pot-maps}), we apply our pipeline to the simulation presented in \citet{Hilbert:2019vca} which they use to compare five different numerical approaches. We re-run the simulation in \textit{gevolution}, starting from a particle snapshot at $z\simeq1.22$ and using a mesh of $3072^3$ grid points, which yields a spatial resolution of $\sim 166.6\, \mathrm{kpc}/h$. This is not quite as good as the maximum resolution reached by the adaptive codes employed in the original comparison, but it is sufficient for our purpose. We then build the light cone for the Newtonian potential and construct the pixel map of $\ka_\mathrm{lin}$ for a source plane at a comoving distance of $2286\, \mathrm{Mpc}/h$, covering the exact same $10 \times 10$ degree patch that was used for the code comparison in \citet{Hilbert:2019vca}.

\begin{figure}
\centering
\includegraphics[width=\columnwidth]{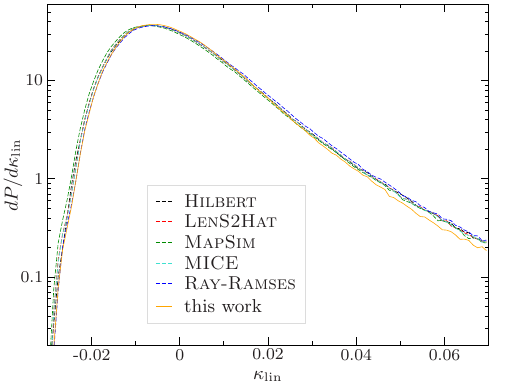}
\vspace{-10pt}
\caption{Comparison of the one-point distributions (smoothed at $1'$) of the convergence obtained for the weak-lensing simulation described in \citet{Hilbert:2019vca} with different numerical methods. Our results have a slight deficit in the tail for large convergence, which is most likely due to the finite resolution of our simulation grid.}
\label{fig:validation}
\end{figure}

After applying a Gaussian smoothing to the scale of one arcminute, we find an excellent agreement of the one-point PDF of $\ka_\mathrm{lin}$ between the different codes, shown in Fig.~\ref{fig:validation}. We therefore conclude that our pixel-based numerical framework is consistent with the current state of the art.

\section*{Data Availability Statement}
The data underlying this article were accessed from the Swiss National
Supercomputing Centre (project ID s710). The derived data generated in
this research will be shared on reasonable request to the corresponding
author.

\bibliographystyle{mnras}
\bibliography{refs} 



\bsp	
\label{lastpage}
\end{document}